%% file: main.tex
\documentclass{article}

\usepackage{arxiv}

\usepackage[utf8]{inputenc} 
\usepackage[T1]{fontenc}    
\usepackage{hyperref}       
\usepackage{url}            
\usepackage{booktabs}       
\usepackage{amsfonts}       
\usepackage{nicefrac}       
\usepackage{microtype}      
\usepackage{lipsum}		
\usepackage{graphicx}
\usepackage[numbers]{natbib}
\usepackage{doi}
\usepackage{amsmath}

\usepackage{algorithm}
\usepackage{algcompatible}
\usepackage{algpseudocode}
\usepackage[%
    prefixes-as-symbols=false,
    ]{siunitx}
\usepackage{tikz}
\usetikzlibrary{shapes,arrows}
\usetikzlibrary{decorations.pathreplacing}
\usetikzlibrary {arrows.meta}
\usetikzlibrary{calc}
\usepackage{pgffor}
\usepackage{tikz-3dplot}
\usepackage{listings}
\usetikzlibrary{decorations.pathreplacing}
\usepackage[braket, qm]{qcircuit}
\usepackage{listings}
\usepackage{cleveref}
\usepackage{lineno}
\usepackage{xcolor}
\usepackage{xspace}
\usepackage{upgreek}
\usepackage{subcaption}
\usepackage{svg}
\usepackage{multirow}
\usepackage{calin}
\usepackage{ifthen}
\usepackage[colorinlistoftodos,prependcaption,textsize=tiny]{todonotes}

\usepackage{pifont}

\title{Quantum Search in Superposed Quantum Lattice Gas Automata and Lattice Boltzmann Systems}

\author{ \href{https://orcid.org/0000-0002-8102-6389}{\includegraphics[scale=0.06]{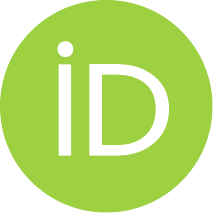}\hspace{1mm}C\u{a}lin A.~Georgescu}\\
	Delft University of Technology\\
	Mekelweg 4, 2628CD, Delft\\
	\texttt{c.a.georgescu@tudelft.nl} \\
	\And
    \href{https://orcid.org/0000-0003-0802-945X}{\includegraphics[scale=0.06]{orcid.pdf}\hspace{1mm}Matthias M\"{o}ller}\\
	Delft University of Technology\\
	Mekelweg 4, 2628CD, Delft\\
	\texttt{m.moller@tudelft.nl} \\
}

\renewcommand{\shorttitle}{Quantum Search in Superposed QLGA and QLBM Systems}

\hypersetup{
pdftitle={Quantum Search in Superposed QLGA and QLBM Systems},
pdfsubject={quant-ph},
pdfauthor={C\u{a}lin A.~Georgescu, Matthias M\"{o}ller},
pdfkeywords={Quantum Computing, Computational Fluid Dynamics, Optimization, Lattice Gas Automata, Lattice Boltzmann Method},
}

\begin{document}
\maketitle
\begin{abstract}
As the scope of Computational Fluid Dynamics (CFD) grows to encompass ever larger problem scales,
so does the interest in whether quantum computing can provide an advantage.
In recent years, Quantum Lattice Gas Automata (QLGA) and Quantum Lattice Boltzmann Methods (QLBM)
have emerged as promising candidates for quantum-native implementations of CFD solvers.
Though the progress in developing QLGA and QLBM algorithms has been significant,
it has largely focused on the development of models rather than applications.
As a result, the zoo of QLGA and QLBM algorithms has grown to target several
equations and to support many extensions, but the practical use of these models
is largely limited to quantum state tomography and observable measurement.
This limitation is crucial in practice, because unless very specific criteria are met,
such measurements may cancel out any potential quantum advantage.
In this paper, we propose an application based on discrete optimization and quantum search,
which circumvents flow field measurement altogether.
We propose methods for simulating many different lattice configurations simultaneously
and describe how the usage of amplitude estimation and quantum search can provide
an asymptotic quantum advantage.
Throughout the paper, we provide detailed complexity analyses of gate-level
implementations of our circuits and consider the benefits and costs of several encodings.
\end{abstract}


\input{1-introduction}
\input{2-preliminaries}

\input{3-parallel-lga}
\input{4-qae-qmf}
\input{6-conclusion}
\input{7-ack}


\bibliographystyle{plainnat}
\bibliography{references}
\end{document}

%% file: 1-introduction.tex
\section{Introduction \label{sec:mp-1-intro}}

Scientific computing has become and indispensable pillar
of the modern scientific and engineering landscapes, spanning
fields that affect nearly all levels of society.
Of the computational sciences, Computational Fluid Dynamics (CFD)
stands out as a particularly versatile tool.
In the aerospace industry, for instance, advances in CFD are crucial
for improving the design of aircraft components and for
lowering the high cost incurred by running experiments \cite{mani2023perspective}.
To realize simulations of the scale and precision required for such applications,
it is common for solvers to utilize thousands of compute nodes
to solve systems with hundreds of millions or billions of unknowns
\cite{mani2023perspective}.
One of the critical challenges that solvers must overcome
to accommodate ever larger industrial demands comes from hardware limitations.
In contrast to the sustained and rapid advancement of hardware capabilities over the last
decades, recent hardware progression has been decelerating.
Faced with the long predicted slowdown of Moore's Law \cite{schaller1997moore}, 
scientists and engineers have turned their attention to alternative
paradigms, including quantum computing \cite{givi2020quantum}.

The first exploratory studies of quantum computing for CFD
date back to the latter half of the 
1990s \cite{meyer1996quantum, yepez1998lattice} and focus on the development
of the first Quantum Lattice Gas Automata (QLGA) systems.
Though several other branches of Quantum CFD methods have since developed
\cite{gaitan2020finding, kyriienko2021solving, demirdjian2022variational, esmaeilifar2024quantum},
QLGA and the related Quantum Lattice Boltzmann Method (QLBM)
have been receiving increasing amounts of interest in recent years.
The QLGA and QLBM research landscape spans many applications, including
transport methods \cite{todorova2020quantum, schalkers2024importance},
the advection-diffusion equation \cite{budinski2021quantum,wawrzyniak2025linearized,wawrzyniak2025dynamic},
the Navier-Stokes equations \cite{budinski2021quantum,steijl2020quantum,itani2023qalb,wang2025quantum},
as well as linearization techniques \cite{itani2022analysis,sanavio2025carleman} and
algorithmic extensions \cite{zamora2025efficient,zamora2025float,fonio2023quantum,fonio2025adaptive,georgescu2025fully,luacuatucs2025surrogate}.
This recent body of work has largely focused on model development,
and has shown that QLGA and QLBM
algorithms are promising candidates for the future development of
quantum CFD solvers.

However, one key challenge that remains unaddressed in this research
is the application of these models.
In much of the literature, models are validated against classical counterparts
by comparing the flow field extracted from the quantum state
inferred by the model.
Though crucial for validation, this method is infeasible for
practical uses of the model, as it requires quantum state tomography (QST)
to sufficiently approximate the quantities encoded in the quantum state.
Clearly, any computational advantage attained by the model up to measurement
is lost in this scenario.
Previous work by \citet{schalkers2024momentum} and \citet{georgescu2025fully}
has formulated observables for the extraction of Quantities of Interest (QoIs)
out of the quantum state for several physical properties, but the efficiency
of these methods is highly dependent on the simulation and only preserves
the advantage in limited scenarios.

In this paper, we propose a novel application of QLGA and QLBM models
that is practically relevant and entirely circumvents the measurement of the flow field.
In particular, we focus on \emph{discrete optimization} scenarios,
in which we are interested in problems where
we aim to find the best candidate solution out of a given input set.
We introduce a novel \emph{parallel time-evolution} extension to
QLGA and QLBM algorithms that enables multiple systems to be simulated simultaneously.
We provide quantum circuits that efficiently accumulate QoIs
in quantum state using light-weight implementations.
To solve the minimization problem, we use established quantum algorithms,
including amplitude estimation \cite{brassard2000quantum}
and the \dhalg~minimum finding algorithm \cite{durr1996quantum}
to obtain a quadratic asymptotic computational
advantage compared to classical counterparts.
Throughout the paper, we give gate-level descriptions of the circuits
we introduce and analyze their complexity under practical assumptions.
Up to the measurement required by the minimum-finding routine,
our algorithm is entirely coherent.
To the best of our knowledge, this marks the first connection
between fields of Quantum CFD and Optimization fields.

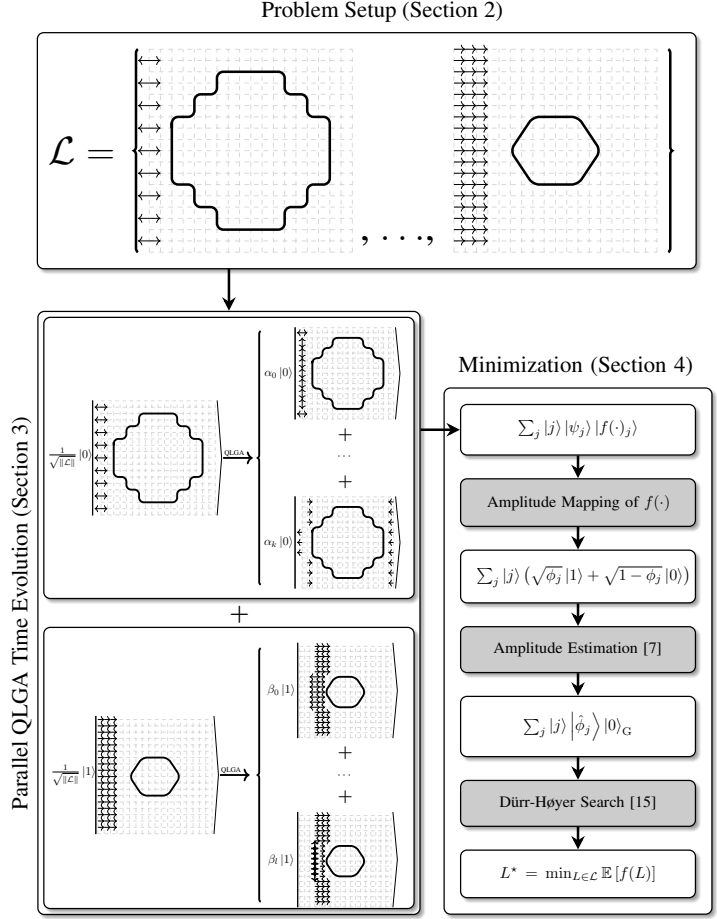
\begin{figure}
    \input{diag-mp-intro-overview.tex}
    
    \caption{Overview of the quantum search algorithm with over superposed QLGA states. \label{fig:mp-1-intro-overview}}
\end{figure}

\Cref{fig:mp-1-intro-overview} provides an end-to-end overview of
our algorithm, which consists of three distinct stages.
The first stage samples several lattice configurations and
formulates the optimization as described in \Cref{sec:mp-2-preliminaries}.
The second stage performs the parallel time-evolution of all sampled systems
with either the QLGA or QLBM algorithms, as detailed in \Cref{sec:mp-3-pqlga}.
Following the time-evolution of the system, the third and final stage
transforms the quantum state such that amplitude estimation and minimum finding
routines can solve the minimization problem with a quadratic asymptotic quantum advantage,
as outlined in \Cref{sec:mp-4-qaeqmf}.
Finally, \Cref{sec:mp-7-conclusion} concludes the paper.

%% file: diag-mp-intro-overview.tex
\usetikzlibrary{shadows.blur}
\usetikzlibrary{backgrounds}
\usetikzlibrary{fit}
\usetikzlibrary{positioning}
\usetikzlibrary{arrows.meta}
\usetikzlibrary{calc}

\centering
\scalebox{0.6}{
\begin{tikzpicture}[scale=0.125]
\begin{scope}[
    xshift=60cm, yshift=-47.5cm,
    node distance=0.5cm,
    box/.style={
        draw, rounded corners, thick,
        blur shadow={shadow blur steps=5, shadow xshift=2pt, shadow yshift=-2pt},
        align=center, inner sep=1em,
        text width=3cm
    }
]
    \node[box, fill=white, text width=4.5cm] (box1) {
        $\sum_j \ket{j}\ket{\psi_j}\ket{f(\cdot)_j}$ 
    };
    \node[box, fill=gray!40, below=of box1, text width=4.5cm] (box2) {
        Amplitude Mapping of $f(\cdot)$
    };
    \node[box, fill=white, below=of box2, text width=4.5cm] (box3) {
        $\sum_j \ket{j}\left(\sqrt{\phi_j}\ket{1} + \sqrt{1-\phi_j}\ket{0} \right)$
    };
    \node[box, fill=gray!40, below=of box3, text width=4.5cm] (box4) {
        Amplitude Estimation \cite{brassard2000quantum}
    };
    \node[box, fill=white, below=of box4, text width=4.5cm] (box5) {
        $\sum_j \ket{j}\ket{\hat{\phi}_j}\ket{0}_{\mathrm{G}}$
    };
    \node[box, fill=gray!40, below=of box5, text width=4.5cm] (box6) {
        \dhalg~Search \cite{durr1996quantum}
    };
	\node[box, fill=white, below=of box6, text width=4.5cm] (box7) {
        $L^\star = \min_{L \in \mathcal{L}} \mathbb{E} \left[ f(L) \right]$
    };

    \foreach \i [evaluate=\i as \j using int(\i+1)] in {1,...,6} {
        \draw[-{Stealth[length=3mm, width=3mm]}, ultra thick] (box\i.south) -- (box\j.north);
    }

    \begin{pgfonlayer}{background}
        \node[
            fit={(box1) (box7)},
            inner sep=1em,
            draw,
            fill=white,
            rounded corners,
            thick,
            blur shadow={shadow blur steps=5, shadow xshift=2pt, shadow yshift=-2pt}
        ] (stage3-box) {};
    \end{pgfonlayer}

    \node[font=\Large,yshift=4em,xshift=1em,text width=6cm] (at=stage3-box.north) {\centering{Minimization (\Cref{sec:mp-4-qaeqmf})}};
\end{scope}

\begin{scope}[yshift=0cm]
	\node (L-label) at (-28, 2) {\Huge $\mathcal{L} = $};
	\draw [decorate,decoration = {brace}, ultra thick] (-18,-16) --  (-18,20);
	
	\draw[
	help lines,
	line width=0.3pt,
	color=gray!30,
	dashed
	] (-16, -16) grid[step={($(2, 2) - (0, 0)$)}] (20, 20);

	\draw[ultra thick, rounded corners] (-12, 7) -- (-12, -4) -- (-8, -4) -- (-8, -8) -- (-4, -8) -- (-4, -12)
	-- (8, -12) -- (8, -8) -- (12, -8) -- (12, -4) -- (16, -4)
	-- (16, 8) -- (12, 8) -- (12, 12) -- (8, 12) -- (8, 16)
	-- (-4, 16) -- (-4, 12) -- (-8, 12) -- (-8, 8) -- (-12, 8) -- (-12, 6);

	\foreach \y in {-14,-10,...,20} {
		\velocityarrow{(-16, \y)}{(-14, \y)};
		\velocityarrow{(-16, \y)}{(-18, \y)};
	}

	\node at (28, -14) {\Huge , $\ldots$,};
	
	\pgfmathsetmacro{\offset}{54}
	\draw[
	help lines,
	line width=0.3pt,
	color=gray!30,
	dashed
	] (\offset-16, -16) grid[step={($(2, 2) - (0, 0)$)}] (\offset+20, 20);

		\foreach \xoffset in {0, 2, 4} {
	\foreach \y in {-14,-10,...,20} {
		\velocityarrow{(\offset-16+\xoffset, \y)}{(\offset-14+\xoffset, \y)};
		\velocityarrow{(\offset-16+\xoffset, \y+2)}{(\offset-14+\xoffset, \y+2)};
		}
	}

	\draw[ultra thick, rounded corners] (\offset+6, 8) -- (\offset-2, 8) -- (\offset-6, 2) -- (\offset-2, -4) -- (\offset+6, -4) -- (\offset+10, 2) -- cycle;
	\draw [decorate,decoration = {brace,mirror}, ultra thick] (76,-16) --  (76,20);
	
	\path (\offset+22, 22) coordinate (top-right-l0);
	\path (-16, -18) coordinate (bottom-left-l0);

	\begin{pgfonlayer}{background}
		\node[
			fit={(L-label) (top-right-l0) (bottom-left-l0) (80,0)},
			draw,
			fill=white,
			rounded corners,
			thick,
			blur shadow={shadow blur steps=5, shadow xshift=2pt, shadow yshift=-2pt}
		] (stage1-box) {};
	\end{pgfonlayer}

    \node[font=\Large,yshift=9.5em,xshift=9em] (at=stage1-box.north) {Problem Setup (\Cref{sec:mp-2-preliminaries})};
\end{scope}


\begin{scope}[xshift=-20cm,yshift=-70cm]
	\begin{scope}[transform canvas={scale=0.75}]
	\node[draw=none, inner sep=0pt] (Lattice) at (0,0) {
		\begin{tikzpicture}[scale=0.125*0.75]
		\begin{scope}[shift={(0,0)}]
			\draw[
				help lines,
				line width=0.3pt,
				color=gray!30,
				dashed
			] (-16, -16) grid[step={($(2, 2) - (0, 0)$)}] (20, 20);

			\draw[ultra thick, rounded corners] (-12, 7) -- (-12, -4) -- (-8, -4) -- (-8, -8) -- (-4, -8) -- (-4, -12)
			-- (8, -12) -- (8, -8) -- (12, -8) -- (12, -4) -- (16, -4)
			-- (16, 8) -- (12, 8) -- (12, 12) -- (8, 12) -- (8, 16)
			-- (-4, 16) -- (-4, 12) -- (-8, 12) -- (-8, 8) -- (-12, 8) -- (-12, 6);

			\foreach \y in {-14,-10,...,20} {
				\velocityarrow{(-16, \y)}{(-14, \y)};
				\velocityarrow{(-16, \y)}{(-18, \y)};
			}
		\end{scope}
		\end{tikzpicture}
	};

	\draw[thick]
	([xshift=-10]Lattice.north west) -- ([xshift=-10]Lattice.south west);
	\draw[thick]
	([xshift=10]Lattice.north east) -- ([xshift=40]Lattice.east)  -- ([xshift=10]Lattice.south east);
	\draw[thick]
	([xshift=-10]Lattice.north west) -- ([xshift=-10]Lattice.south west) node at ([xshift=-160]Lattice.west) {$\frac{1}{\sqrt{\| \mathcal{L} \|}}\ket{0}$};

	\draw[->, ultra thick, black] ([xshift=50]Lattice.east)-- ([xshift=250]Lattice.east) node at ([xshift=125,yshift=30]Lattice.east) {\tiny{QLGA}};
	\draw [decorate,decoration = {brace}, ultra thick] (25,-30) --  (25,30);

	\node[draw=none, inner sep=0pt] (Lattice1) at (45,20) {
		\begin{tikzpicture}[scale=0.125*0.6]
		\begin{scope}[shift={(0,0)}]
			\draw[
				help lines,
				line width=0.3pt,
				color=gray!30,
				dashed
			] (-16, -16) grid[step={($(2, 2) - (0, 0)$)}] (20, 20);

			\draw[ultra thick, rounded corners] (-12, 7) -- (-12, -4) -- (-8, -4) -- (-8, -8) -- (-4, -8) -- (-4, -12)
			-- (8, -12) -- (8, -8) -- (12, -8) -- (12, -4) -- (16, -4)
			-- (16, 8) -- (12, 8) -- (12, 12) -- (8, 12) -- (8, 16)
			-- (-4, 16) -- (-4, 12) -- (-8, 12) -- (-8, 8) -- (-12, 8) -- (-12, 6);

			\foreach \y in {-14,18} {
				\velocityarrow{(-16, \y)}{(-14, \y)};
				\velocityarrow{(-16, \y)}{(-18, \y)};
			}

			\foreach \y in{-10,-6,-2,2,6,10,14}{
				\velocityarrow{(-16, \y)}{(-16, \y+2)};
				\velocityarrow{(-16, \y)}{(-16, \y-2)};
			}
		\end{scope}
		\end{tikzpicture}
	};

	\draw[thick]
	([xshift=-10]Lattice1.north west) -- ([xshift=-10]Lattice1.south west);
	\draw[thick]
	([xshift=10]Lattice1.north east) -- ([xshift=40]Lattice1.east)  -- ([xshift=10]Lattice1.south east);
		\draw[thick]
	([xshift=-10]Lattice1.north west) -- ([xshift=-10]Lattice1.south west) node at ([xshift=-120]Lattice1.west) {$\alpha_0\ket{0}$};
	\draw[] node at ([yshift=-100]Lattice1.south) {$\Large\boldsymbol{+}$};
	\draw[] node at ([yshift=-250]Lattice1.south) {$\cdots$};
		\node[draw=none, inner sep=0pt] (Lattice2) at (45,-20) {
		\begin{tikzpicture}[scale=0.125*0.6]
		\begin{scope}[shift={(0,0)}]
			\draw[
				help lines,
				line width=0.3pt,
				color=gray!30,
				dashed
			] (-16, -16) grid[step={($(2, 2) - (0, 0)$)}] (20, 20);

			\draw[ultra thick, rounded corners] (-12, 7) -- (-12, -4) -- (-8, -4) -- (-8, -8) -- (-4, -8) -- (-4, -12)
			-- (8, -12) -- (8, -8) -- (12, -8) -- (12, -4) -- (16, -4)
			-- (16, 8) -- (12, 8) -- (12, 12) -- (8, 12) -- (8, 16)
			-- (-4, 16) -- (-4, 12) -- (-8, 12) -- (-8, 8) -- (-12, 8) -- (-12, 6);

			\foreach \y in {-14,-10,-6,-2,2,6,10,14,18} {
				\velocityarrow{(20, \y)}{(18, \y)};
			}

			\foreach \y in {-14,-10,-6,10,14,18} {
				\velocityarrow{(-14, \y)}{(-12, \y)};
			}
			
			\foreach \y in {-2,2,6} {
				\velocityarrow{(-16, \y)}{(-18, \y)};
			}
		\end{scope}
		\end{tikzpicture}
	};
	\draw[] node at ([yshift=100]Lattice2.north) {$\Large\boldsymbol{+}$};

	\draw[thick]
	([xshift=-10]Lattice2.north west) -- ([xshift=-10]Lattice2.south west);
	\draw[thick]
	([xshift=10]Lattice2.north east) -- ([xshift=40]Lattice2.east)  -- ([xshift=10]Lattice2.south east);
	\draw[thick]
	([xshift=-10]Lattice2.north west) -- ([xshift=-10]Lattice2.south west) node at ([xshift=-120]Lattice2.west) {$\alpha_k\ket{0}$};

\end{scope}
\path (60*0.75+5, 32*0.75+17.5) coordinate (top-right-l1);
\path (-25*0.75+5, -32*0.75+17.5) coordinate (bottom-left-l1);
\begin{pgfonlayer}{background}
	\node[
		fit={(top-right-l1) (bottom-left-l1)},
		draw,
		fill=white,
		rounded corners,
		thick,
		blur shadow={shadow blur steps=5, shadow xshift=2pt, shadow yshift=-2pt}
	] (system1-box) {};
\end{pgfonlayer}

\draw[] node at (20,-12.75+2.5) {$\Large\boldsymbol{+}$};
\end{scope}

\begin{scope}[xshift=-20cm,yshift=-145cm]
	\begin{scope}[transform canvas={scale=0.75}]
	\node[draw=none, inner sep=0pt] (Lattice) at (0,0) {
		\begin{tikzpicture}[scale=0.125*0.75]
		\begin{scope}[shift={(0,0)}]
			\draw[
			help lines,
			line width=0.3pt,
			color=gray!30,
			dashed
			] (-16, -16) grid[step={($(2, 2) - (0, 0)$)}] (20, 20);
			\draw[ultra thick, rounded corners] (+6, 8) -- (-2, 8) -- (-6, 2) -- (-2, -4) -- (+6, -4) -- (+10, 2) -- cycle;

			\foreach \xoffset in {0, 2, 4} {
				\foreach \y in {-14,-10,...,20} {
					\velocityarrow{(-16+\xoffset, \y)}{(-14+\xoffset, \y)};
					\velocityarrow{(-16+\xoffset, \y+2)}{(-14+\xoffset, \y+2)};
				}
			}

		\end{scope}
		\end{tikzpicture}
	};

	\draw[thick]
	([xshift=-10]Lattice.north west) -- ([xshift=-10]Lattice.south west);
	\draw[thick]
	([xshift=10]Lattice.north east) -- ([xshift=40]Lattice.east)  -- ([xshift=10]Lattice.south east);
	\draw[thick]
	([xshift=-10]Lattice.north west) -- ([xshift=-10]Lattice.south west) node at ([xshift=-160]Lattice.west) {$\frac{1}{\sqrt{\| \mathcal{L} \|}}\ket{1}$};

	\draw[->, ultra thick, black] ([xshift=50]Lattice.east)-- ([xshift=250]Lattice.east) node at ([xshift=125,yshift=30]Lattice.east) {\tiny{QLGA}};
	\draw [decorate,decoration = {brace}, ultra thick] (25,-30) --  (25,30);

	\node[draw=none, inner sep=0pt] (Lattice1) at (45,20) {
		\begin{tikzpicture}[scale=0.125*0.6]
		\begin{scope}[shift={(0,0)}]
			\draw[
			help lines,
			line width=0.3pt,
			color=gray!30,
			dashed
			] (-16, -16) grid[step={($(2, 2) - (0, 0)$)}] (20, 20);
			
			\foreach \xoffset in {6, 8} {
				\foreach \y in {-14,-10,...,20} {
					\velocityarrow{(-16+\xoffset, \y)}{(-14+\xoffset, \y)};
					\velocityarrow{(-16+\xoffset, \y+2)}{(-14+\xoffset, \y+2)};
				}
			}

			\foreach \y in {-14,-10,-6,10,14,18} {
				\ifthenelse{\y = -6}{}{\velocityarrow{(-16+10, \y+2)}{(-14+10, \y+2)};}

				\velocityarrow{(-16+10, \y)}{(-14+10, \y)};
			}

			\foreach \y in {-6, -2, 2, 6} {
				\ifthenelse{\y = -6}{}{\velocityarrow{(-14+4, \y)}{(-16+4, \y)};}
				
				\velocityarrow{(-14+4, \y+2)}{(-16+4, \y+2)};
				
			}

			\draw[ultra thick, rounded corners] (+6, 8) -- (-2, 8) -- (-6, 2) -- (-2, -4) -- (+6, -4) -- (+10, 2) -- cycle;

		\end{scope}
		\end{tikzpicture}
	};

	\draw[thick]
	([xshift=-10]Lattice1.north west) -- ([xshift=-10]Lattice1.south west);
	\draw[thick]
	([xshift=10]Lattice1.north east) -- ([xshift=40]Lattice1.east)  -- ([xshift=10]Lattice1.south east);
		\draw[thick]
	([xshift=-10]Lattice1.north west) -- ([xshift=-10]Lattice1.south west) node at ([xshift=-120]Lattice1.west) {$\beta_0\ket{1}$};
	\draw[] node at ([yshift=-100]Lattice1.south) {$\Large\boldsymbol{+}$};
	\draw[] node at ([yshift=-250]Lattice1.south) {$\cdots$};
	
		\node[draw=none, inner sep=0pt] (Lattice2) at (45,-20) {
		\begin{tikzpicture}[scale=0.125*0.6]
		\begin{scope}[shift={(0,0)}]
			\draw[
			help lines,
			line width=0.3pt,
			color=gray!30,
			dashed
			] (-16, -16) grid[step={($(2, 2) - (0, 0)$)}] (20, 20);
			
			\foreach \xoffset in {8} {
				\foreach \y in {-14,-10,...,20} {
					\velocityarrow{(-16+\xoffset, \y)}{(-14+\xoffset, \y)};
					\velocityarrow{(-16+\xoffset, \y+2)}{(-14+\xoffset, \y+2)};
				}
			}

			\foreach \xoffset in {6, 10} {
				\foreach \y in {-14,-10,-6,10,14,18} {
					\ifthenelse{\y = -6}{}{\velocityarrow{(-16+\xoffset, \y+2)}{(-14+\xoffset, \y+2)};}

					\velocityarrow{(-16+\xoffset, \y)}{(-14+\xoffset, \y)};
				}
			}

			\foreach \y in {-4, -2, 0, 2, 4, 6, 8} {
				\ifthenelse{\y = -6}{}{
					\velocityarrow{(-14+4, \y)}{(-14+4, \y+2)};
					\velocityarrow{(-14+4, \y)}{(-14+4, \y-2)};
				}
				
				\velocityarrow{(-14+4, \y)}{(-14+4, \y+2)};
				\velocityarrow{(-14+4, \y)}{(-14+4, \y-2)};
				
			}

			\draw[ultra thick, rounded corners] (+6, 8) -- (-2, 8) -- (-6, 2) -- (-2, -4) -- (+6, -4) -- (+10, 2) -- cycle;

		\end{scope}
		\end{tikzpicture}
	};
	\draw[] node at ([yshift=100]Lattice2.north) {$\Large\boldsymbol{+}$};

	\draw[thick]
	([xshift=-10]Lattice2.north west) -- ([xshift=-10]Lattice2.south west);
	\draw[thick]
	([xshift=10]Lattice2.north east) -- ([xshift=40]Lattice2.east)  -- ([xshift=10]Lattice2.south east);
	\draw[thick]
	([xshift=-10]Lattice2.north west) -- ([xshift=-10]Lattice2.south west) node at ([xshift=-120]Lattice2.west) {$\beta_l\ket{1}$};

\end{scope}
\path (60*0.75+5, 32*0.75+37.5) coordinate (top-right-l2);
\path (-25*0.75+5, -32*0.75+37.5) coordinate (bottom-left-l2);
\begin{pgfonlayer}{background}
	\node[
		fit={(top-right-l2) (bottom-left-l2)},
		draw,
		fill=white,
		rounded corners,
		thick,
		blur shadow={shadow blur steps=5, shadow xshift=2pt, shadow yshift=-2pt}
	] (system2-box) {};
\end{pgfonlayer}
\end{scope}

\begin{pgfonlayer}{background}
    \node[
        fit={(system1-box) (system2-box)},
        draw,
        fill=white,
        rounded corners,
        thick,
        blur shadow={shadow blur steps=5, shadow xshift=2pt, shadow yshift=-2pt}
    ] (systems-box) {};
\end{pgfonlayer}
\begin{pgfonlayer}{background}
	\node[
		fit={(top-right-l1) (bottom-left-l1)},
		draw,
		fill=white,
		rounded corners,
		thick,
		blur shadow={shadow blur steps=5, shadow xshift=2pt, shadow yshift=-2pt}
	] (system1-box) {};
\end{pgfonlayer}
\begin{pgfonlayer}{background}
	\node[
		fit={(top-right-l2) (bottom-left-l2)},
		draw,
		fill=white,
		rounded corners,
		thick,
		blur shadow={shadow blur steps=5, shadow xshift=2pt, shadow yshift=-2pt}
	] (system1-box) {};
\end{pgfonlayer}

\node[rotate=90, anchor=south, font=\Large] at (systems-box.west) {Parallel QLGA Time Evolution (Section 3)};

\draw[-{Stealth[length=3mm, width=3mm]}, ultra thick]  ($(systems-box.north) + (0, 7.5)$) -- (systems-box.north);
\draw[-{Stealth[length=3mm, width=3mm]}, ultra thick]  ($(box1.west) + (-7, 0)$) -- ($(box1.west)$);
\end{tikzpicture}
}

%% file: 2-preliminaries.tex
\section{Preliminaries \label{sec:mp-2-preliminaries}}

This section provides the necessary background and definitions that our algorithm is built on top of.
For the remainder of the paper, we frame the description of our methods
for linear encoding of the QLGA algorithm, as it provides the
both the most challenges and the most opportunities.
We address typical QLBM encodings and complexities in \Cref{subsec:mp-3-4-alternatives},
and note fundamental differences wherever they arise.
We first describe the linear QLGA encoding we use throughout the rest of the paper,
before formalizing the problem statement.
Finally, we provide a succinct overview of the unstructured search, amplitude estimation,
and minimum finding quantum algorithms.

\subsection{Quantum Lattice Gas Automata and Quantum Lattice Boltzmann Methods \label{subsec:mp-2-2-qlga}}

The origins of Cellular Automata (CA) can be traced back to the seminal works
on foundational models of computation by John von Neumann and Arthur Burks around
the middle of the 20$^{\text{th}}$ century \cite{v1966theory}.
In a general sense, CA models consist of structured arrangements of discrete cells,
which are assigned one of a finite number of discrete states,
and evolve according to a set of rules that are applied
simultaneously to all cells \cite{wolf2004lattice}.
The work of \citet{hardy1973time} in 1973 is the first
application of CA in the domain of computational physics and marks the emergence
of the Lattice Gas Cellular Automata (LGA) subfamily of CA.
This first LGA model introduced a 2-dimensional grid structure
of cells arranged in a square structure with links to their nearest neighbors
in both directions of either dimension.
Particle behavior is governed by particle transport (streaming) and
particle-particle interactions (collisions) that change
the course of particles upon intersection.
The more complex hexagonal grid LGA model of \citet{frisch1986lattice} later superseded
its predecessor as its larger symmetry group
can be shown to lead to the incompressible Navier-Stokes
equations in the hydrodynamic limit \cite{wolf2004lattice}.
These models are known in the literature as
the HPP \cite{hardy1973time} and FHP \cite{frisch1986lattice} models, respectively.

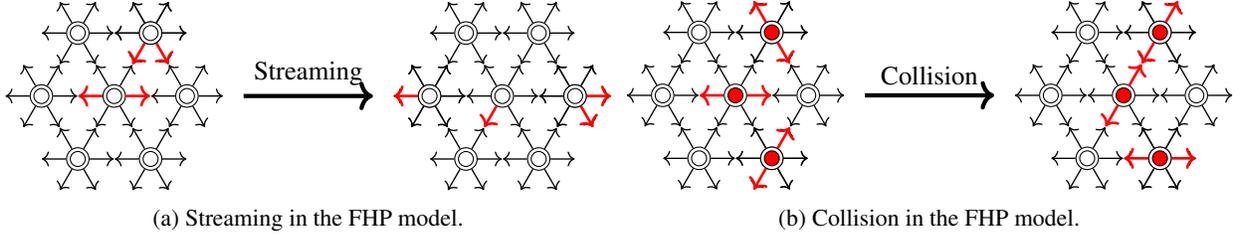
\begin{figure}
    \centering
    \hfill
    \subcaptionbox{Streaming in the FHP model.\label{fig:mp-2-fhp-streaming}}{\input{diag-mp-fhp-streaming.tex}}
    \hfill
    \subcaptionbox{Collision in the FHP model.\label{fig:mp-2-fhp-collision}}{\input{diag-mp-fhp-collision.tex}}
    \hfill
    \caption{Example of lattice states, streaming, and collision in the FHP model \cite{frisch1986lattice}. \label{fig:mp-2-fhp-example}}
\end{figure}

One key distinction between CA and LGA is the physical interpretation assigned
to states of the grid.
Generally, an LGA gridpoint consists of several distinct \emph{velocity channels},
which act as means of propagating information between gridpoints.
The LGA grid is populated by discrete particles, which have pre-determined properties
and are indistinguishable from one another.
The state of an LGA gridpoint is determined by whether each of its velocity channels
is occupied by a particle.
Since particles are indistinguishable, this state can be modelled
as a bitstring whose length is equal to the number of velocity channels.
The rules that govern the behavior of LGA  -- streaming and collision --
are visually depicted in \Cref{fig:mp-2-fhp-example} for the FHP discretization.
\Cref{fig:mp-2-fhp-streaming} shows how particles (highlighted in red) stream
along velocity channels, while \Cref{fig:mp-2-fhp-collision} portrays how collision
stochastically redistributes particle occupancies (at the highlighted gridpoints) such
that mass and momentum are conserved.
This straightforward physical discretization leads naturally to the computation of
physical properties of the system, such as mass and momentum-density \cite{wolf2004lattice}.
A general description of QLGA for arbitrary
velocity discretizations is given in \citet{georgescu2025fully}.

Around a decade after the development of the FHP model,
interest in the quantum realization  of LGA began to surge.
The first Quantum Lattice Gas Automata models were developed by 
Meyer \cite{meyer1996quantum, meyer1997quantumb},
Boghosian \cite{boghosian1997quantum,boghosian1998simulating},
and Yepez \cite{yepez1998lattice,yepez1998quantum,yepez2001quantum,yepez2002quantum}
and established the feasibility and limitations of implementing the stream-collide
semantics of LGA on quantum computers.
However, shortly after the turn of the century,
as classical CFD methods greatly benefitted from the unprecedented growth fueled
by the rapid improvement of computer hardware,
the largely theoretical QLGA research experienced a decline in interest.
Following nearly two decades of stagnation,
the resurgence in QLGA research occurred in the late 2010s as
researchers again 
turned their attention to potential CFD Applications of quantum computing.
\citet{love2019quantum} introduced quantum circuits for the
collision and propagation steps for several 1- and 2-dimensional
QLGA models.
\citet{fonio2023quantum}, \citet{kocherla2024fully}, and \citet{georgescu2025fully}
introduce extensions to the initialization, streaming, collision, and measurement
steps of the QLGA loop.
\citet{fonio2023quantum} and \citet{zamora2025efficient} utilize a compressed encoding,
which more efficiently represents the state of a lattice at the cost of
performing restarts following every time step.
\citet{zamora2025float} and \citet{fonio2025adaptive} introduce quantum
algorithms based on variations of the Integer LGA -- a model which
replaces the boolean interpretation of particles with numerical values.

In this work, we focus our analysis on the \emph{linear encoding}
discussed in \cite{love2019quantum,kocherla2024fully,georgescu2025fully},
which maps bits and qubits bijectively.
Though this encoding offers no direct memory advantage, it allows for
the simulation of arbitrarily many coherent time steps and
simultaneous representation of exponentially
many lattice states \cite{schalkers2024importance,fonio2023quantum, wang2025quantum}.
The representational power of this encoding is important, as QoIs
are generally computed in terms of \emph{ensemble averages}
over many stochastic outcomes \cite{wolf2004lattice}.
As we cover in the following sections, the linear encoding
provides further computational advantages that are lost
when information is compressed into fewer qubits.

\paragraph{Relation to Random and Quantum Walks.} The boolean-particle LGA algorithm
can be interpreted as a random walk over a weighted directed graph $G = (V, E)$, where
the vertex set $V$ is the set of all attainable lattice configurations,
and the edge set $E$ is determined by the collision semantics of the system.
Applying the typical collision operators, which preserve macroscopic quantities
such as mass and momentum, one can define the stochastic physical
semantics in terms of transition probabilities of a Markov chain
traversal of $G$.
For a number of gridpoints $N_g$ and a discretization of $q$ velocity
channels, the size of the state space is $\mathcal{O}(2^{qN_g})$.
Analogously, the QLGA algorithm can be interpreted as a discrete-time quantum walk,
where branching occurs every LGA step by means of the collision step, while the
streaming step maps each state to a neighboring vertex.
For simulating a large number of time steps --
and therefore a sizeable proportion of the Markov state space --
the bijective linear encoding becomes necessary to simultaneously
accommodate all outcomes.
For thorough analyses of quantum walks and the broader field of quantum cellular automata,
we refer the reader to references \cite{venegas2012quantum} and \cite{farrelly2020review}, respectively.

\subsubsection{Quantum Lattice Boltzmann Methods}

The Lattice Boltzmann Method \cite{succi2001lattice,kruger2017lattice} has historically
succeeded LGA as a computational approach for fluid flow problems.
The core addition of the LBM is the introduction of a global distribution
function that replaces the boolean particle discretization of its predecessor.
This leads to significant noise reduction, as the LBM
inherently tracks the averaged behavior of fluid at the mesoscopic scale, rather
than the individual trajectory of each particle as in LGA.
To accommodate this smoother discretization, the collision step of the LBM
departs from the granular particle redistributions of the LGA models
and is instead defined as a change in the local distribution associated with each gridpoint.
The Bhatnagar-Gross-Krook (BGK) \cite{bhatnagar1954model} collision operator
is one of the simplest and most common implementation of the collision operator,
which can still recover the bulk Navier-Stokes properties, and can be seen
as a relaxation of the local distribution towards equilibrium.

Interest in the quantum analog of the LBM began to surge in
the early 2020s with works such as those of Budinski \cite{budinski2021quantum,ljubomir2022quantum}
and \citet{steijl2020quantum}.
Though QLBM research has progressed significantly,
two of the main open challenges are the modelling of nonlinearity and dissipation.
The BGK \cite{bhatnagar1954model} collision operator and
more sophisticated counterparts that solve the Navier-Stokes equations
all require the computation
of quadratic terms used in irreversible updates of the distribution function.
Since this operation is not directly implementable in the common quantum encodings
used in the literature, much of the QLBM research has focused on developing
suitable approximations of the collision process.
This includes the usage of the Linear Combination of Unitaries
\cite{childs2012hamiltonian} by \citet{ljubomir2022quantum},
approximating the collision process by Carleman Linearization \cite{itani2022analysis,sanavio2024lattice},
applying a classical correction step to the quantum state \cite{wang2025quantum},
and learning a linear approximation of the collision \cite{luacuatucs2025surrogate,itani2025towards}.
To utilize the methods described in this paper, we only require that
the QLBM solver is coherent, and assume that it uses an amplitude-based encoding, \ie,
that the value of the distribution function
is encoded in the amplitude of each basis state.
Two recently proposed approaches that fall into this category include
the heat-transfer QLBM solver of \citet{jawetz2025quantum} and the advection-diffusion
solver of \citet{nagel2025quantum}.

\paragraph{Nomenclature and Complexity Analaysis.}
Throughout this paper, we provide circuit-level
complexity analyses of the proposed algorithms.
To do this, we assume an all-to-all connectivity of logical qubits,
and we measure complexity in terms of 1- and 2- qubit
gate decompositions.
If no efficient decomposition is known for a particular gate,
we assume its decomposition is exponentially expensive in the number
of qubits it spans.
We use the notations $N_{q, (\cdot)}$ and $N_{t, (\cdot)}$
to denote the number of qubits and time steps associated with
a particular component of the algorithm, respectively.
Finally we denote the application of a gate $U$ with $p$
controls as $\mathrm{C}^pU$.

\subsection{Problem Statement\label{subsec:mp-2-1-problem}}

We describe our algorithms as solvers for an optimization problem that is formulated as follows.
Let $\mathcal{L}$ be the set of lattice configurations that share the same spatial,
temporal, and velocity discretizations, but that differ in either (i) initial conditions
or (ii) boundary conditions.
For each lattice $L \in \mathcal{L}$, let $f(L, \Omega, t) \in \mathbb{R}$ be the value of some scalar
quantity of interest (QoI) computed over the physical
region of space $\Omega$ of $L$ at time $t$.
We use the shorthand notation 

\begin{equation}
    f(L, \Omega) = \sum_{t \in \mathcal{N}_{t, acc}} f(L, \Omega, t)
\end{equation}
to denote the cumulative value of the quantity of interest over a specific
set of discrete time steps $\mathcal{N}_{t, acc}$.
Finally, let $\mathrm{LGA}(L, N_t)$ be the probability distribution
that arises after the time-evolution of lattice configuration
of the lattice $L$ for $N_t$ time steps by the LGA algorithm.
In this work, we are concerned with finding the lattice configuration

\begin{equation}
	L^\star = \min_{L \in \mathcal{L}} \mathbb{E} _{x \sim \mathrm{LGA}(L, N_t)}\left[ f(x, \Omega) \right],
\label{eq:mp-1-1-problem}
\end{equation}

for a pre-defined lattice configuration $L$, QoI $f$ and region $\Omega$.
That is, the lattice for which the QoI is minimal
in the expectation
over the ensemble average that emerges from nondeterminstic collision\footnote{Maximization is also trivially realizable by adjusting the threshold and Grover oracle in the \dhalg~step.}.
Practically, the QoI encoded in the function $f$ could be the drag coefficient
of an airfoil, or the average pressure over a region of interest.
The set $\mathcal{L}$ could consist of several airfoil designs subject to the same
initial conditions, 
or distinct set of initial conditions applied to the same geometry.

The QLBM counterpart of the problem statement is similar,
but contains two key differences.
First, the ensemble averaging that QLGA relies on is no longer
required in the QLBM setting.
Second, we restrict ourselves to the computation of the QoI
for a single time step.
This is not a theoretical limitation, but rather
a practical consideration that stems from the
amplitude-based encoding that is typical of QLBM methods.
Estimating the mean value of an amplitude over multiple time steps
is possible, but would utilize circuits whose complexity
would dominate the typical QLBM time step.

\subsection{Amplitude Amplification, Estimation, and Quantum Search \label{subsec:mp-2-3-grover}}

This section briefly reviews the key concepts behind the Quantum Amplitude Amplification (QAA),
Quantum Amplitude Estimation (QAE), and Quantum Search algorithms.
We begin with a short description of Grover's celebrated algorithm for unstructured database search.

\subsubsection{Unstructured Database Search and Grover's Algorithm \label{subsubsec:mp-2-3-1-grover}}

The unstructured database search problem consists of (efficiently) searching a structureless
table $T$ of $N$ entries where $T[j] \in \{ 0, 1 \}~\forall~j \in \{0, ..., N-1\}$.
The goal is to output an index $j$ such that $T[j] = 1$, if one exists.
Throughout this work, we are concerned with the case where $t$ such entries
exist, for an unknown number $t \geq 1$.
The complexity of an algorithm for this problem is typically measured in
the number of times the algorithm \emph{queries} the table $T$, \ie,
the number of times an item of $T$ is accessed.
Classically, algorithms that solve the unstructured database search problem
can be reduced to sampling a hypergeometric distribution without replacement,
yielding an expected number of queries that is $\mathcal{O}(N/t)$.
One of the most celebrated results in quantum computing stems from
Grover's algorithm \cite{grover1996fast,grover1997quantum} that has been shown to solve the unstructured database
search problem using quadratically fewer queries.

To understand Grover's algorithm, we can partition the table $T$ into two sets

\begin{equation}
    \begin{cases}
        A_0 = \{ j~|~T[j] = 0 \},\\
        A_1 = \{ j~|~T[j] = 1 \}
    \end{cases}
\end{equation}

with $A_1$ the set of solutions and $A_0$ the set of non-solutions.
Grover's algorithm begins by initializing a register in the uniform superposition

\begin{equation}
    \begin{split}
        \ket{u} & = \sum_{j=0}^{N-1} \frac{1}{\sqrt{N}} \ket{j} = \sum_{j \in A_1} k_0\ket{j} + \sum_{j \in A_0} l_0\ket{j} = \sqrt{\frac{\| A_1 \|}{N}}\ket{A_1} + \sqrt{\frac{\| A_0 \|}{N}}\ket{A_0},
    \end{split}
    \label{eq:mp-2-3-good-and-bad-states}
\end{equation}

with $k_0 = l_0 = 1/\sqrt{N}$.
Grover's algorithm can be interpreted as a series of
two-dimensional rotations in the space spanned by
$\ket{A_1}$ and $\ket{A_0}$ (often referred to as the "good" and "bad" states),
where each rotation is composed of reflections about
the $\ket{A_1}$ and $\ket{u}$ states, respectively.
This operation is known as the Grover iterator

\begin{equation}
    \mathcal{G} = \underbrace{\left( \mathrm{H}^{\otimes n} \mathrm{S}_{\text{0}} \mathrm{H}^{\otimes n}\right)}_{= 2 \ket{u}\bra{u} - \mathbb{I}}S_T,
    \label{eq:mp-2-grover-iterator}
\end{equation}

where $S_T$ is a \emph{phase query gate} that acts as $S_T\ket{x} = -1^{T[x]}\ket{x}$, and
$\mathrm{S}_0 = 2 \ket{0}^{\otimes n}\bra{0}^{\otimes n} - \mathbb{I}$
is the reflection about the $\ket{0}^{\otimes n}$ state.
After $j$ executions of the Grover iteration, the coefficients can be
geometrically interpreted as

\begin{equation}
    \begin{cases}
        k_j = \frac{1}{\sqrt{t}} \sin((2j+1)\theta),\\
        l_j = \frac{1}{\sqrt{N-t}} \cos((2j+1)\theta),
    \end{cases}
    \label{eq:mp-2-sin-amplitudes}
\end{equation}



where the angle $\theta$ satisfies $\sin^2 \theta = t/N$.
Intuitively, the goal is to select an integer number of iterations
such that the rotations move the state close to the good state $\ket{A_1}$,
or, equivalently, such that $l_j$ is close to 0.
For an in-depth analysis of the case where the number of solutions is unknown,
we refer the reader to the works of \citet{boyer1998tight} and \citet{brassard2000quantum}, where it is shown
that the number of queries required for this case is $\mathcal{O}(\sqrt{N/t})$.

\subsubsection{Quantum Amplitude Amplification and Estimation}

Quantum Amplitude Amplification (QAA) \cite{brassard2000quantum} is a generalization
of Grover's original unstructured database search algorithm \cite{grover1996fast, grover1997quantum}.
Applied to the same setting described in \Cref{subsubsec:mp-2-3-1-grover},
QAA is tasked with increasing the probability of measuring the "good" state 
$\ket{A_1}$ by means of the Grover iterator.
In this interpretation, the QAA algorithm can be understood as increasing
the amplitude of $\ket{A_1}$ by roughly a constant at each iteration,
leading to a quadratic improvement in the probability of measuring this
state at the end of the execution \cite{brassard2000quantum}.

Quantum Amplitude Estimation (QAE) is an extension of QAA, in which we
are interested in estimating the probability of measuring the 
good state $a = \langle A_1 | A_1 \rangle$ as given in \Cref{eq:mp-2-3-good-and-bad-states}, 
or, equivalently, estimating $\alpha$ in a quantum state state of the form $\sqrt{\alpha}\ket{1} + \sqrt{1-\alpha}\ket{0}$.
The first formulation of QAE was proposed by \citet{brassard2000quantum},
and uses several applications of the iterator in \Cref{eq:mp-2-grover-iterator}
controlled on the state of an ancillary register mapped to the Fourier basis.
QAE leverages the sinusoidal form of the amplitudes in \Cref{eq:mp-2-sin-amplitudes}
to estimate the angle $\theta \in [0, \uppi/2]$ and therefore obtains an estimate
$\hat{\alpha} = \sin^2(\uppi y/N)$ for the state $y$
measured in the ancillary register \cite{brassard2000quantum}.
This method -- applying multiple controlled Grover iterators followed by the inverse QFT --
closely resembles Quantum Phase Estimation (QPE)
and has become known as the \emph{canonical} QAE in the literature.
The canonical QAE requires a query complexity of $\mathcal{O}(\epsilon^{-1})$
for an additive error $\epsilon$,
which is in alignment with the theoretical bounds derived by \citet{nayak1999quantum},
and a quadratic improvement over the classical $\mathcal{O}(\epsilon^{-2})$.
The Grover operator used in this algorithm is given by 

\begin{equation}
    \mathcal{Q} = \mathcal{A}\mathrm{S}_0\mathcal{A}^\dagger S_T,
    \label{eq:mp-2-qae-iterator}
\end{equation}

with $S_0$ and $S_T$ defined as in the \Cref{eq:mp-2-grover-iterator},
while $\mathcal{A}$ denotes the coherent quantum algorithm that acts as

\begin{equation}
    \mathcal{A}\ket{0}^{\otimes n}\ket{0} = \sqrt{\alpha}\ket{A_1}\ket{1} + \sqrt{1-\alpha}\ket{A_0}\ket{0}.
    \label{eq:mp-2-oracle-form}
\end{equation}

In recent years, several approaches have emerged, that seek to reduce
the cost of QAE by reducing its reliance on the QPE building blocks,
while retaining its query advantage.
\citet{suzuki2020amplitude} introduced a variant of QAE, which samples
several applications of the Grover iterator and approximates $\alpha$ classically
by means of maximum likelihood estimation.
\citet{wie2019simpler} introduces an alternative QAE algorithm that relies on
applications of controlled Grover iterators with increasingly many controls,
without requiring the use of the $\qft$, and instead relying on the Hadamard test.
\citet{aaronson2020quantum} provide an analysis of another QAE variant
that resembles binary search over the number of applications of Grover iterators.
\citet{grinko2021iterative} analyze yet another entirely Grover-based
QAE algorithm and show an empirical improvement over previous counterparts.
While all of these algorithms remove the requirement on the $\qft$ that
the canonical QAE uses, they all require measurements to estimate
the target probability $\alpha$.
However, we utilize QAE as an intermediate step for estimating the mean of our
QLGA distribution, which we follow with a minimum-finding routine.
Therefore, we require the QAE algorithm to be coherent, and for
this reason rely on the canonical QAE instead of the more modern counterparts.

%% file: diag-mp-fhp-streaming.tex
\begin{tikzpicture}[scale=0.215]

\dtwoqsixcolor{(0,0)}{0.1cm}{2}{0}{1.5}{white}{red}{black}{black}{red}{black}{black}

\foreach \i in {0,...,5} {
    \pgfmathsetmacro{\angle}{60*\i}
    \pgfmathsetmacro{\x}{4.5*cos(\angle)}
    \pgfmathsetmacro{\y}{4.5*sin(\angle)}
    \dtwoqsix{(\x,\y)}{0.1cm}{2}{\i+1}{1.5}{white}
}

\pgfmathsetmacro{\angle}{60*1}
\pgfmathsetmacro{\x}{4.5*cos(\angle)}
\pgfmathsetmacro{\y}{4.5*sin(\angle)}
\dtwoqsixcolor{(\x,\y)}{0.1cm}{2}{0}{1.5}{white}{black}{black}{black}{black}{red}{red}

\draw[->, ultra thick] (8, 0) -- (16, 0) node [above, midway] {Streaming};

\dtwoqsixcolor{(24,0)}{0.1cm}{2}{0}{1.5}{white}{black}{black}{black}{black}{red}{black}
\foreach \i in {0,...,5} {
    \pgfmathsetmacro{\angle}{60*\i}
    \pgfmathsetmacro{\x}{4.5*cos(\angle)}
    \pgfmathsetmacro{\y}{4.5*sin(\angle)}
    \dtwoqsix{(24+\x,\y)}{0.1cm}{2}{\i+1}{1.5}{white}
}

\pgfmathsetmacro{\angle}{60*6}
\pgfmathsetmacro{\x}{4.5*cos(\angle)}
\dtwoqsixcolor{(\x + 24,0)}{0.1cm}{2}{0}{1.5}{white}{red}{black}{black}{black}{black}{red}

\pgfmathsetmacro{\angle}{60*3}
\pgfmathsetmacro{\x}{4.5*cos(\angle)}
\dtwoqsixcolor{(\x + 24,0)}{0.1cm}{2}{0}{1.5}{white}{black}{black}{black}{red}{black}{black}
\end{tikzpicture}

%% file: diag-mp-fhp-collision.tex
\begin{tikzpicture}[scale=0.215]

\dtwoqsixcolor{(0,0)}{0.1cm}{2}{0}{1.5}{red}{red}{black}{black}{red}{black}{black}

\foreach \i in {0,...,5} {
    \pgfmathsetmacro{\angle}{60*\i}
    \pgfmathsetmacro{\x}{4.5*cos(\angle)}
    \pgfmathsetmacro{\y}{4.5*sin(\angle)}
    \dtwoqsix{(\x,\y)}{0.1cm}{2}{\i+1}{1.5}{white}
}

\pgfmathsetmacro{\angle}{60*1}
\pgfmathsetmacro{\x}{4.5*cos(\angle)}
\pgfmathsetmacro{\y}{4.5*sin(\angle)}
\dtwoqsixcolor{(\x,\y)}{0.1cm}{2}{0}{1.5}{red}{black}{black}{red}{black}{black}{red}

\pgfmathsetmacro{\angle}{60*5}
\pgfmathsetmacro{\x}{4.5*cos(\angle)}
\pgfmathsetmacro{\y}{4.5*sin(\angle)}
\dtwoqsixcolor{(\x,\y)}{0.1cm}{2}{0}{1.5}{red}{black}{red}{black}{black}{red}{black}

\draw[->, ultra thick] (8, 0) -- (16, 0) node [above, midway] {Collision};

\dtwoqsixcolor{(24,0)}{0.1cm}{2}{0}{1.5}{red}{black}{red}{black}{black}{red}{black}
\foreach \i in {0,...,5} {
    \pgfmathsetmacro{\angle}{60*\i}
    \pgfmathsetmacro{\x}{4.5*cos(\angle)}
    \pgfmathsetmacro{\y}{4.5*sin(\angle)}
    \dtwoqsix{(24+\x,\y)}{0.1cm}{2}{\i+1}{1.5}{white}
}

\pgfmathsetmacro{\angle}{60*1}
\pgfmathsetmacro{\x}{4.5*cos(\angle)}
\pgfmathsetmacro{\y}{4.5*sin(\angle)}
\dtwoqsixcolor{(24+\x,\y)}{0.1cm}{2}{0}{1.5}{red}{black}{red}{black}{black}{red}{black}

\pgfmathsetmacro{\angle}{60*5}
\pgfmathsetmacro{\x}{4.5*cos(\angle)}
\pgfmathsetmacro{\y}{4.5*sin(\angle)}
\dtwoqsixcolor{(24+\x,\y)}{0.1cm}{2}{0}{1.5}{red}{red}{black}{black}{red}{black}{black}
\end{tikzpicture}

%% file: 3-parallel-lga.tex
\section{Parallel QLGA Time Evolution and Quantity Accumulation \label{sec:mp-3-pqlga}}

This section describes the parallel QLGA time evolution algorithm
and its building blocks.
We first detail the quantum register setup and analyze the complexity of
known building blocks in the linear qubit encoding.
We follow with a description of parallel time evolution and efficient quantity accumulation
and conclude the section with a discussion of alternative encoding methods. 

\subsection{Quantum Register Setup and Base Operations}

Throughout this section, we assume that we are addressing distinct lattice configurations
that have $N_g$ gridpoints and $q$ velocity channels each,
following the commonplace \dq{d}{q} notation.
We further assume that we are optimizing over a finite
set $\mathcal{L}$, with $\| \mathcal{L} \|$ lattice distinct configurations.
The number of qubits required to implement the linear encoding for a single lattice is
simply $qN_g$, which we refer to as the base register $B$.

Within the base register, a single layer of $\mathcal{O}(N_g)$ single-qubit gates is required
to implement uniformly distributed initial conditions.
The $\mathcal{O}(1)$ depth of this operation is one of the computational advantages
of the linear encoding over its logarithmically compressed counterparts.
The streaming step can be implemented by means of $q$ parallel
applications of $N_g - 1$ swap gates per streaming direction, each
of depth $\lceil \log_2 N_g \rceil$ \cite{schalkers2024importance,georgescu2025fully}.
Another advantage of the the linear encoding is that the complexities
related to efficiently applying typical bounce-back and specular reflection boundary conditions
around solid geometries are largely eliminated.
Since the boundary conditions remain consistent across all instances
of the superposed LGA states of a single configuration, 
and since the grid information is
not compressed, there is no need to control boundary condition
applications on any ancillary state.
As a result, the imposition of boundary conditions discussed in \cite{schalkers2024efficient} and 
\cite{georgescu2025fully} on geometries with a
perimeter spanning $N_{bc}$ gridpoints requires $\mathcal{O}(qN_{bc})$ swap gates
that act completely in parallel, with depth $1$.
The cost and application of boundary conditions is consistent with the
description of \cite{georgescu2025fully}.

In addition to the base register, we require 5 additional registers for our algorithm.
First is a \emph{marker} register $M$ that we use to distinguish between the
$\| \mathcal{L} \|$ lattice configurations.
We shall henceforth assume a maximally compressed representation of this data
by encoding the markers into $\lceil \log_2 \| \mathcal{L} \| \rceil$ qubits, and assigning
each basis state to a configuration.
We discuss an alternative encoding and its tradeoffs in \Cref{subsec:mp-3-4-alternatives}.
Second is a \emph{data accumulation} register $D$ in which the algorithm tracks the
total value of the QoI throughout the algorithm.
For a problem in which we are interested in tracking
a scalar quantity over a region $\Omega$ for $N_{t, acc}$ time steps,
we generally require at most
$\lceil \log_2\left( N_{t, acc} \cdot \| \Omega \| \cdot (q + 1) \right) \rceil$ qubits.
We provide more details about this choice in \Cref{subsec:mp-3-3-accumulation}.
Third, we require an additional, single-qubit "coin" register $C$, which we use to transform
our state into a form that is suitable for the application of QAE and unstructured search.
A fourth register $E$ of size $e$ is assigned for running the amplitude estimation routine,
and its size $e$ register is left as a choice to the user.
Finally, we use $1$ more qubit $G$ to distinguish good states during Grover search.
The total number of qubits our algorithm uses is therefore
\begin{equation}
    N_q = \underbrace{qN_g}_{\text{Base qubits}} + \overbrace{\lceil \log_2 \| \mathcal{L} \| \rceil}^{\text{Marker qubits}} + \underbrace{\lceil \log_2\overbrace{\left(N_{t, acc} \cdot \| \Omega \| \cdot (q + 1) \right)}^{\text{Maximum value~} F_{\mathrm{max}}} \rceil}_{\text{$N_{q,acc}$ Accumulation qubits}} + \overbrace{e}^{\text{QAE qubits}} + \underbrace{2.}_{\text{Coin and Grover qubits}}
    \label{eq:mp-3-qubit-number}
\end{equation}

For the application to QLBM algorithms, the base register qubit count
is typically reduced to $\lceil \log_2 q \rceil + \lceil \log_2 N_g \rceil$
by virtue of the amplitude-based encoding.
The accumulation register may be omitted in both cases, should the
method be applied to a single time step.

\subsection{Parallel QLGA Time Evolution of Configurations}

The parallel-configuration QLGA algorithm closely resembles the core
QLGA loop, with two key exceptions.
The base operations of streaming and collision are used identically
as in the QLGA algorithm and retain the same semantics throughout.
Where the parallel QLGA algorithm differs is in its application of initial and boundary conditions.
Unlike streaming and collisions, these operations cannot be applied
uniformly across configurations, since the their semantics are the only differentiation
between lattice configurations.
It is for this purpose that we employ the marker register $M$ to distinguish between lattice.
A straightforward way to implement the different configuration-specific semantics is to
control the application of initial and boundary conditions on the state of the marker.

To realize this implementation,
we set the state of the register to the uniform superposition
$\ket{u}_\mathrm{M} = 1/\sqrt{\| \mathcal{L} \|} \sum_{j=0}^{\| \mathcal{L} \| - 1} \ket{j}$, and assign
each basis state $\ket{j}$ to represent a lattice.
Using this information, we can simply iterate through each basis state
to control the configuration-specific semantics.
Since both (uncontrolled) uniform initial conditions and
reflection boundary conditions can be applied in parallel with a depth of 1,
and there are $\lceil \log_2 \| \mathcal{L} \| \rceil$ control qubits,
one can implement the controlled counterparts of the operations of a lattice $L$
using $N_{bc, L}$ $\CPX{\lceil \log_2 \| \mathcal{L} \| \rceil}$ gates.

Applying sequential boundary condition imposition
for all geometry across all configurations and using the gate
decomposition introduced by \citet{barenco1995elementary} leads
to a circuit with size and depth of $\mathcal{O}(N_{bc, L_{\text{max}}} \cdot \| \mathcal{L} \| \cdot \log^2 \| \mathcal{L} \|)$,
where $L_{\text{max}}$ denotes the lattice with the geometry that spans the largest number of gridpoints.
The first two terms stem from the serialized imposition across all gridpoints of all lattices,
while the quadratic logarithmic term is due to the $\mathrm{CX}$ decomposition.
Importantly, the streaming and collision circuits used in QLGA are
entirely agnostic of the marker register.
This, in turn, means the complexity of these steps
remains unaffected by the additions of our algorithm.

\paragraph{Exploiting overlap.} One can reduce the cost of enforcing configuration-specific semantics
if several configurations share common features.
For a particular feature of the configuration of $L$, let $\mathcal{V}_L$ denote the finite set of
velocity channels involved in enforcing the semantics of the feature.
For two different configurations, $L_j$ and $L_k$, the intersection
$\mathcal{V}_{L_j} \cap \mathcal{V}_{L_k} \neq \emptyset$
provides an opportunity to enforce the conditions of both configurations using fewer control qubits.
One can do this in two steps: first permuting the basis states $\ket{j}$ and $\ket{k}$
into the $\ket{1}^{\otimes \lceil \log_2 \| \mathcal{L} \| \rceil-1}\ket{0}$ and $\ket{1}^{\otimes \lceil \log_2 \| \mathcal{L} \| \rceil-1}\ket{1}$ states, and second
controlling the operation only on the first $\lceil \log_2 \| \mathcal{L} \| \rceil-1$ qubits.
In general, for an overlapping region of $k$ lattices such that $\cap_{j=1}^k \mathcal{V}_{L_{j-1}} \neq \emptyset$,
one can reduce the number of controls of the semantics of $\lfloor \log_2 k \rfloor$ of these lattices to
$\lceil \log_2 \| \mathcal{L} \| \rceil - \lfloor \log_2 k \rfloor$.
Asymptotically, this method is in fact worse than the non-overlapping counterpart,
since permutations generally require quadratically
many one-and two qubit gates to implement \cite{herbert2024almost}.
Clearly, addressing all exponentially many possible intersections is extremely inefficient.
However, intelligently exploiting scenarios in which the intersections of different lattices
are sizeable (\ie, assessing how small differences in airfoil design affect drag at the leading edge of an airplane)
and appear in many configurations can lead to significant practical improvements.

We note that in instances where the number of lattices we simulate is not a power of 2,
some marker qubit states remain unused.
Following the initialization of the system, the quantum state would
therefore contain some basis states $\ket{0}^{\otimes qN_g}_{\mathrm{B}}\ket{k}_{\mathrm{M}}$ for the basis
states $k$ that are unassigned.
Following the common QLGA semantics, neither collision nor streaming can alter such states
as they would otherwise violate mass and momentum conservation.
We therefore do not include such cases in the rest of the analyses in this paper.

\paragraph{Example.}
To help provide a more intuitive understanding of how superposed semantics can be implemented in
practice, we provide an example of 3 distinct lattices, as depicted in \Cref{fig:mp-3-example-lattice-1},
\Cref{fig:mp-3-example-lattice-2}, and \Cref{fig:mp-3-example-lattice-3}, first in terms
of initial conditions and then in terms of boundary treatment.
In all 3 lattices, the arrows indicate the initial state of the flow field, while the
black outlines delimitate the perimeters of three different boundary-conditioned objects.
Our algorithm for this instance requires $qN_g$ qubits for the QLGA loop,
and only two additional qubits to mark the three lattice configurations.

\begin{figure}
    \centering
    \hfill
    \subcaptionbox{Configuration $\mathrm{L_1} \mapsto \ket{00}_\mathrm{M}$.\label{fig:mp-3-example-lattice-1}}{\input{diag-mp-parallel-lattice-1.tex}}
    \hfill
    \subcaptionbox{Configuration $\mathrm{L_2} \mapsto \ket{01}_\mathrm{M}$.\label{fig:mp-3-example-lattice-2}}{\input{diag-mp-parallel-lattice-2.tex}}
    \hfill
    \subcaptionbox{Configuration $\mathrm{L_3} \mapsto \ket{10}_\mathrm{M}$.\label{fig:mp-3-example-lattice-3}}{\input{diag-mp-parallel-lattice-3.tex}}
    \hfill\strut
    \subcaptionbox{Schematic of superposed initial condition quantum circuit.\label{fig:mp-3-example-schematic-init}}{\input{circ-mp-parallel-qlga-schematic-init.tex}}
    \hfill\strut
    \subcaptionbox{Schematic of LGA time step  quantum circuit with superposed boundary condition imposition.\label{fig:mp-3-example-schematic-loop}}{\input{circ-mp-parallel-qlga-schematic-loop.tex}}
    \caption{Example of the parallel QLGA circuit for a set of 3 lattices with intersecting initial and boundary conditions.}
\end{figure}

The initial state of the system is therefore $\ket{0}^{\otimes qN_g}_\mathrm{B}\ket{++}_{\mathrm{M}}$.
To impose the initial conditions effectively, we first notice that while the initial
state of $\mathrm{L_2}$ does not overlap with the two other lattices,
there \emph{is} a significant overlap between $\mathrm{L_1}$ and $\mathrm{L_3}$.
Using this information, we can devise a quantum circuit that implements these semantics
with sequential steps that address $\mathrm{L_2}$ controlled on both marker qubits,
the shared semantics $\mathcal{V}_\mathrm{L_1} \cap \mathcal{V}_\mathrm{L_3}$ controlled on only one of the marker qubits,
and the remainder $\mathcal{V}_\mathrm{L_1} - \mathcal{V}_\mathrm{L_3}$ and $\mathcal{V}_\mathrm{L_3} - \mathcal{V}_\mathrm{L_1}$, respectively.
The schematic implementing this routine is depicted in \Cref{fig:mp-3-example-schematic-init}.

Once the initial conditions are set, the usual QLGA streaming and collision steps can
commence completely agnostically of the marker qubits, as depicted by the first
2 operations of \Cref{fig:mp-3-example-schematic-loop}.
Next, analyzing the properties of the boundary conditions of the three systems
we observe that there is a shared segment between the three objects, namely the
left wall of the square of $\mathrm{L_2}$.
We can enforce the boundary conditions of this wall using the standard $\mathcal{O}(1)$ depth
circuit afforded by the linear encoding completely in parallel from the rest of the
boundary conditions.
Following this step, we can proceed by iterating through the remainder geometries sequentially,
accounting for the shared segments that should not be repeated.
This circuit can then be repeated for all time steps without modification.

\subsection{Quantity Accumulation \label{subsec:mp-3-3-accumulation}}

Though the QLGA and QLBM literature has introduced several
techniques for computation of physical quantities out of the quantum state
without the use of quantum state tomography \cite{schalkers2024momentum, fonio2023quantum, georgescu2025qlbm},
these techniques have all focused on extracting the information out of the quantum
state at a given time step by means of measurement.
In contrast, we require a circuit that coherently \emph{accumulates} the required quantity across 
multiple time steps.
Before describing the design of our circuit, we first analyze the properties of the quantities
that one might want to query in the QLGA algorithm.

Within the LGA discretization, physical quantities such as mass, momentum, and forces
can all be computed based on the boolean particle occupancies \cite{wolf2004lattice}.
This in turn means that even more complex quantities, such as drag and lift coefficients,
can be computed linearly and coherently from the quantum state.
Moreover, the boolean weights of particles make it such that physical quantities
can be expressed proportionally to the Hamming weight of the bitstrings
encoding the gridpoints within the region of interest.
The only exception to this rule is that it is common for discretizations to
include a so-called \emph{rest particle}, which is typically assigned twice the mass of regular particles.
Accounting for this, we require a circuit that computes the function

\begin{equation}
    f(\Omega) \propto \sum_{x\in \Omega} \sum_{j \in {\mathcal{V}}} \alpha_j x_j
\end{equation}

with $\Omega$ the region of interest, $x$ the gridpoints within this section, $\mathcal{V}$ the
set of velocity channels that contribute to the QoI we compute, and $\alpha_j$ the
mass of the velocity channel in lattice units.
For commonplace discretizations, $\alpha_j \in \{1, 2\}$.
We can easily derive an upper bound for the number of qubits required to store
this value throughout our computation by leveraging the fact that the value of this
sum is at most $\| \mathcal{V} \|+ 1$ (accounting for the rest particle)
for each gridpoint and each time step, which implies that we require at most

\begin{equation}
    N_{q, acc} = \left\lceil \log_2 \left( N_{t, acc} \cdot \| \Omega \| \cdot (q + 1) \right)  \right\rceil
\end{equation}

qubits to retain this value without overflow.
To efficiently accumulate this quantity,
we introduce a Modified Hamming Weight Adder (MHWA)
based on the design of the Draper Adder \cite{draper2000addition}.

The Draper Adder was originally designed
to perform addition between two registers, such that the such that
$\mathrm{U}\ket{x}\ket{y} = \ket{x}\ket{x+y}$, by means of the Quantum Fourier Transform (QFT)
and (controlled) phase gates \cite{draper2000addition}.
Our modified adder instead performs the operation

\begin{equation}
    U_{\mathrm{MHWA}}\ket{x}\ket{y} = \ket{x}\ket{y + \sum_{j=0}^{\| x \| - 1} \alpha_j x_j}.
    \label{eq:mp-3-mhwa}
\end{equation}

To implement this operation, we can fix the parameters of the controlled phase gates that
are applied to the Fourier-transformed $\ket{y}$ state such that the operation effectively
implements $U_\mathrm{P}\ket{y} = \ket{y + 1}$ controlled on the state of a single qubit of $\ket{x}$.
This is the same technique as used in the controlled streaming operator described by \citet{schalkers2024efficient},
and the parameters of the phase gates can be easily adjusted to accommodate the 
weight of the rest particle at no additional complexity cost.
We can then iterate through all qubits of $\ket{x}$ repeating this operation to obtain the exact form
described in \Cref{eq:mp-3-mhwa}.

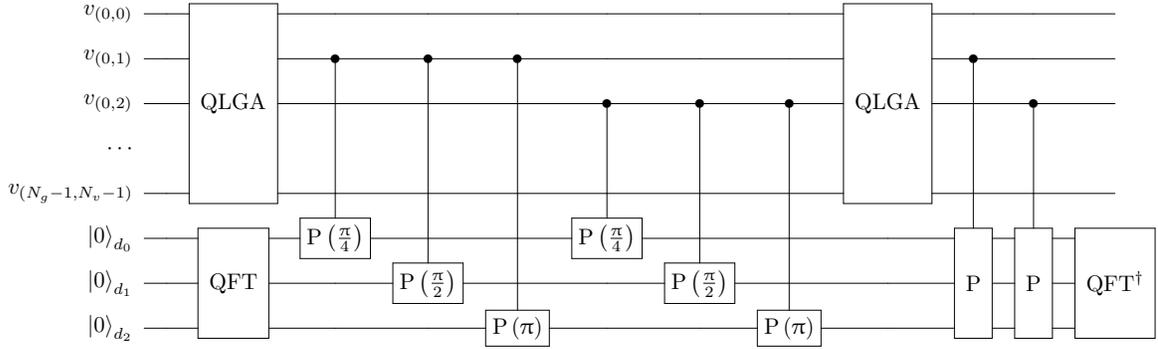
\begin{figure}
    \input{circ-mp-mhwa-schematic.tex}
    \caption{Schematic of Modified Hamming Weight Adder quantum circuit.\label{fig:mp-3-mhwa-schematic}}
\end{figure}

The integration of this procedure within the parallel-configuration QLGA loop is straightforward,
and can be implemented as depicted in \Cref{fig:mp-3-mhwa-schematic}.
We begin by initializing the data register $D$ to the QFT of the vacuum state and perform the QLGA
loop in parallel.
To accumulate the quantities, we perform iterative controlled phase gates at the end of each time step,
where the controls are the qubits that correspond to the region of interest $\Omega$.
Note that unless the QLGA loop has reached its end, there is no need to perform the inverse QFT,
as following accumulation steps require the data qubits to be in Fourier space.
As such, for each time step, we execute only the controlled phase gates
(depicted succinctly as the $\mathrm{P}$ gates in \Cref{fig:mp-3-mhwa-schematic}) and map the qubits back
to the computational basis at the end of the computation.
Furthermore, this implementation is entirely independent of the marker register,
as we are comparing the same QoI across all configurations.
Assuming $N_{q,acc}$ is the number of accumulation qubits as described by \Cref{eq:mp-3-qubit-number},
the MHWA circuit can be realized using exactly $\| \Omega \| N_{q,acc}$ controlled phase gates,
in addition to the $\mathcal{O}\left( N_{q,acc}^2\right)$ gates required
to implement the $\qft$ blocks.
The depth of the controlled phase array can be reduced to $\mathcal{O}\left(\| \Omega \| + N_{q,acc}\right)$,
as each qubit of the accumulation register is addressed independently.
Following the application of $N_t$ time steps consisting of interleaving QLGA and accumulation steps,
we obtain a quantum state of the form

\begin{equation}
    \ket{\psi_{\mathrm{LGA}}} = \sum_{j} \ket{\psi_{L_j}}_{\mathrm{B}}\ket{f(L_j, \Omega)}_{\mathrm{D}}\ket{j}_{\mathrm{M}}\ket{0}_{\mathrm{C}}\ket{0}^{\otimes e}_{\mathrm{E}}\ket{0}_{\mathrm{G}}.
    \label{eq:mp-3-quantum-state-after-acc}
\end{equation}

\subsection{Alternative Encodings\label{subsec:mp-3-4-alternatives}}

Before considering the optimization part of our algorithm, we briefly consider
the effects that different encoding choices entail.
We first analyze the consequences of encoding choices of the
marker register, before assessing the application of our methods
to algorithms that utilize different encodings in the base register.

\paragraph{Marker register encodings.}
Thus far, we assumed a maximally compressed encoding
of the marker register due to its qubit efficiency.
However, this method requires heavily controlled
applications in instances in which configurations share little overlap.
An alternative marker register data structure that alleviates
this limitation is the \emph{one-hot encoding}.
In the one-hot encoding, we use $\| \mathcal{L} \|$
marker qubits, and assign each configuration a state

\begin{equation}
    \ket{\boldsymbol{e}_j} = \ket{0}^{\otimes j}\ket{1}\ket{0}^{\otimes \| \mathcal{L} \| - j - 1},
\end{equation}

for $0 \leq j \leq \| \mathcal{L} \| - 1$.
This encoding makes it such that each configuration can be identified
by a single qubit, instead of by entire marker register.
While the qubit requirement increases from $\mathcal{O}(\log_2 \| \mathcal{L} \|)$
in the maximally compressed scenario to $\mathcal{O}(\| \mathcal{L} \|)$
for the one-hot alternative, the semantics of the linear encoding
QLGA boundary conditions allow for a significant gate count and depth reduction.
Since boundary conditions of disjoint geometries can be enforced
fully in parallel, and since each configuration is identified
by a single qubit, we can therefore apply disjoint
boundary condition circuits on arbitrarily many lattices
by adding a single control to the appropriate swap gates.
While the number of gates still scales with the perimeter of the
geometry, the depth of this circuit is $\mathcal{O}(1)$
due to disjoint controls and targets.

Which of the two register encoding strategies is superior in practice
is highly dependent on the application and the resources available.
For highly disjointed geometries and hardware with large
numbers of qubits, the one-hot encoding preserves the parallel
application of boundary conditions, which is a
significant advantage of the linear QLGA encoding.
For hardware that is limited in the number of qubits,
but where the computational budget
is more permissive and where overlap can be strategically exploited,
the maximally compressed encoding might become preferable,
if the additional cost associated with the decomposition of multi-controlled
$\mathrm{X}$ gates is admissible.

\paragraph{Base register encodings.}
Thus far we have analyzed our algorithm in the linear encoding of
the base register due to its expressivity and computational
advantages.
However, many QLGA and QLBM algorithms explore the
amplitude-based encoding, thanks to its logarithmic
memory compression of the grid register
\cite{budinski2021quantum, ljubomir2022quantum,schalkers2024efficient,steijl2020quantum,wawrzyniak2025linearized}.
While our marker encoding and boundary conditions schemes
can be readily applied with consistent qubit requirements
and computational complexity analyses, QoI handling requires more consideration.
In amplitude-based encodings,
the quantity accumulation method described previously
does not apply, as the physical components that contribute to
the QoIs are tied to amplitudes, rather than Hamming weights.
Previous work by \citet{schalkers2024momentum} and \citet{georgescu2025fully}
has, however, introduced quantum circuits that accumulate
the amplitude spread over physical regions of space onto
a single ancillary qubit, with particularly
efficient implementations for regions that can be bounded
by axis-aligned and diagonal boundaries.
While these methods differ from our implementation, and can only be
directly utilized for one time step, they nonetheless construct
a quantum state that is exactly of the form we need, as we explore in
\Cref{subsec:mp-4-1-amplitude-projection}.

Though computationally expensive,
addressing the multiple time step scenario is also possible by applying
multiple controlled state evolution operators as is the case
in, for instance, quantum phase estimation.
We note that accumulating the quantity over multiple time
steps is not a necessity for all practical applications.
Instances in which QoIs are computed once
the flow has reached a steady state lend themselves
particularly well to the single-time step case.

%% file: diag-mp-parallel-lattice-1.tex
\centering 
\begin{tikzpicture}[scale=0.1]
	\draw[
help lines,
line width=0.3pt,
color=gray!30,
dashed
] (-20, -20) grid[step={($(2, 2) - (0, 0)$)}] (24, 24);

\draw[ultra thick, rounded corners] (-12, 7) -- (-12, -4) -- (-8, -4) -- (-8, -8) -- (-4, -8) -- (-4, -12)
-- (8, -12) -- (8, -8) -- (12, -8) -- (12, -4) -- (16, -4)
-- (16, 8) -- (12, 8) -- (12, 12) -- (8, 12) -- (8, 16)
-- (-4, 16) -- (-4, 12) -- (-8, 12) -- (-8, 8) -- (-12, 8) -- (-12, 6);

\foreach \y in {-18,-14,...,22} {
    \velocityarrow{(-18, \y)}{(-16, \y)};
    \velocityarrow{(-20, \y)}{(-18, \y)};
}
\end{tikzpicture}

%% file: diag-mp-parallel-lattice-2.tex
\centering 
\begin{tikzpicture}[scale=0.1]
	\draw[
help lines,
line width=0.3pt,
color=gray!30,
dashed
] (-20, -20) grid[step={($(2, 2) - (0, 0)$)}] (24, 24);

\draw[ultra thick, rounded corners] (-12, 7) -- (-12, -4) -- (0, -4) -- (0, 8) -- (-12, 8) -- (-12, 6);

\foreach \y in {-18,-14,...,22} {
    \velocityarrow{(-16, \y)}{(-18, \y)};
    \velocityarrow{(-18, \y)}{(-20, \y)};
}
\end{tikzpicture}

%% file: diag-mp-parallel-lattice-3.tex
\centering 
\begin{tikzpicture}[scale=0.1]
	\draw[
help lines,
line width=0.3pt,
color=gray!30,
dashed
] (-20, -20) grid[step={($(2, 2) - (0, 0)$)}] (24, 24);

\draw[ultra thick, rounded corners] (-12, 15) -- (-12, -12) -- (0, -12) -- (0, 16) -- (-12, 16) -- (-12, 6);

\foreach \y in {-18,-14,...,22} {
    \velocityarrow{(-18, \y)}{(-16, \y)};
    \velocityarrow{(-18, \y)}{(-18, \y+2)};
    \velocityarrow{(-18, \y)}{(-18, \y-2)};
}
\end{tikzpicture}

%% file: circ-mp-parallel-qlga-schematic-init.tex
\centering
\scalebox{1}{
\Qcircuit @C=1.0em @R=0.2em @!R { \\
\nghost{{v}_{(0, 0)} :  } & \lstick{{v}_{(0, 0)} } & \qw& \qw & \qw & \multigate{7}{\mathrm{X}_{\mathrm{L_2}}}& \qw &\qw & \qw & \multigate{7}{\mathrm{X}_{\mathrm{L_1 \cap L_3}}} & \multigate{7}{\mathrm{X}_{\mathrm{L_3 - L_1}}} & \qw & \multigate{7}{\mathrm{X}_{\mathrm{L_1 - L_3}}} \\
\nghost{{v}_{(0, 1)} :  } & \lstick{{v}_{(0, 1)} } & \qw& \qw & \qw & \ghost{\mathrm{X}_{\mathrm{L_2}}}& \qw & \qw &\qw & \ghost{\mathrm{X}_{\mathrm{L_1 \cap L_3}}} & \ghost{\mathrm{X}_{\mathrm{L_3 - L_1}}}& \qw& \ghost{\mathrm{X}_{\mathrm{L_1 - L_3}}}& \qw\\
\nghost{{v}_{(0, 2)} :  } & \lstick{{v}_{(0, 2)} } & \qw& \qw & \qw& \ghost{\mathrm{X}_{\mathrm{L_2}}}& \qw & \qw &\qw & \ghost{\mathrm{X}_{\mathrm{L_1 \cap L_3}}} & \ghost{\mathrm{X}_{\mathrm{L_3 - L_1}}}& \qw& \ghost{\mathrm{X}_{\mathrm{L_1 - L_3}}}& \qw\\
\nghost{{v}_{(0, 3)} :  } & \lstick{\cdots }\\
\nghost{{v}_{(0, q - 1)} :  } & \lstick{{v}_{(0, q - 1)} } & \qw& \qw & \qw& \ghost{\mathrm{X}_{\mathrm{L_2}}}& \qw & \qw &\qw & \ghost{\mathrm{X}_{\mathrm{L_1 \cap L_3}}} & \ghost{\mathrm{X}_{\mathrm{L_3 - L_1}}}& \qw& \ghost{\mathrm{X}_{\mathrm{L_1 - L_3}}}& \qw\\
\nghost{{v}_{(1, 0)} :  } & \lstick{{v}_{(1, 0)} } & \qw& \qw & \qw& \ghost{\mathrm{X}_{\mathrm{L_2}}}& \qw & \qw &\qw & \ghost{\mathrm{X}_{\mathrm{L_1 \cap L_3}}} & \ghost{\mathrm{X}_{\mathrm{L_3 - L_1}}}& \qw& \ghost{\mathrm{X}_{\mathrm{L_1 - L_3}}}& \qw\\
\nghost{{v}_{(1, 1)} :  } & \lstick{\cdots }\\
\nghost{{v}_{(N_g - 1, q - 1)} :  } & \lstick{{v}_{(N_g - 1, q - 1)} } & \qw& \qw & \qw& \ghost{\mathrm{X}_{\mathrm{L_2}}}& \qw &\qw & \qw & \ghost{\mathrm{X}_{\mathrm{L_1 \cap L_3}}} & \ghost{\mathrm{X}_{\mathrm{L_3 - L_1}}}& \qw& \ghost{\mathrm{X}_{\mathrm{L_1 - L_3}}}& \qw\\
\nghost{{m}_{0} :  } & \lstick{{m}_{0} } & \qw & \gate{\mathrm{H}} & \gate{\mathrm{X}} & \ctrl{-1} & \gate{\mathrm{X}} & \qw & \qw & \qw & \ctrl{-1} & \gate{\mathrm{X}} & \ctrl{-1}&\gate{\mathrm{X}}\\
\nghost{{m}_{1} :  } & \lstick{{m}_{1} } & \qw & \gate{\mathrm{H}} & \qw & \ctrl{-1} & \qw & \qw & \gate{\mathrm{X}} & \ctrl{-2} & \ctrl{-2} & \qw & \ctrl{-2}&\gate{\mathrm{X}}\\
  \\}}

%% file: circ-mp-parallel-qlga-schematic-loop.tex
\centering
\scalebox{1}{
\Qcircuit @C=1.0em @R=0.2em @!R { \\
\nghost{{v}_{(0, 0)} :  } & \lstick{{v}_{(0, 0)} } & \qw &\multigate{7}{\mathrm{U_{coll}^{\otimes N_g}}}&\multigate{7}{\mathrm{U_{str}}}&\multigate{7}{\mathrm{BC_{L_1 \cap L_2 \cap L_3}}} &\multigate{7}{\mathrm{BC_{L_1}}} & \qw & \multigate{7}{\mathrm{BC_{L_2}}} & \qw & \multigate{7}{\mathrm{BC_{L_3}}}&\qw&\qw\\
\nghost{{v}_{(0, 1)} :  } & \lstick{{v}_{(0, 1)} } & \qw  & \ghost{\mathrm{U_{coll}^{\otimes N_g}}}&\ghost{\mathrm{U_{str}}}&\ghost{\mathrm{BC_{L_1 \cap L_2 \cap L_3}}}&\ghost{\mathrm{BC_{L_1}}}&\qw&\ghost{\mathrm{BC_{L_2}}}&\qw&\ghost{\mathrm{BC_{L_3}}}&\qw&\qw\\
\nghost{{v}_{(0, 2)} :  } & \lstick{{v}_{(0, 2)} } & \qw  & \ghost{\mathrm{U_{coll}^{\otimes N_g}}}&\ghost{\mathrm{U_{str}}}&\ghost{\mathrm{BC_{L_1 \cap L_2 \cap L_3}}}&\ghost{\mathrm{BC_{L_1}}}&\qw&\ghost{\mathrm{BC_{L_2}}}&\qw&\ghost{\mathrm{BC_{L_3}}}&\qw&\qw\\
\nghost{{v}_{(0, 3)} :  } & \lstick{\cdots }\\
\nghost{{v}_{(0, q - 1)} :  } & \lstick{{v}_{(0, q - 1)} } & \qw & \ghost{\mathrm{U_{coll}^{\otimes N_g}}}&\ghost{\mathrm{U_{str}}}&\ghost{\mathrm{BC_{L_1 \cap L_2 \cap L_3}}}&\ghost{\mathrm{BC_{L_1}}}&\qw&\ghost{\mathrm{BC_{L_2}}}&\qw&\ghost{\mathrm{BC_{L_3}}}&\qw&\qw\\
\nghost{{v}_{(1, 0)} :  } & \lstick{{v}_{(1, 0)} } & \qw  & \ghost{\mathrm{U_{coll}^{\otimes N_g}}}&\ghost{\mathrm{U_{str}}}&\ghost{\mathrm{BC_{L_1 \cap L_2 \cap L_3}}}&\ghost{\mathrm{BC_{L_1}}}&\qw&\ghost{\mathrm{BC_{L_2}}}&\qw&\ghost{\mathrm{BC_{L_3}}}&\qw&\qw\\
\nghost{{v}_{(1, 1)} :  } & \lstick{\cdots }\\
\nghost{{v}_{(N_g - 1, q - 1)} :  } & \lstick{{v}_{(N_g - 1, q - 1)} }  & \qw & \ghost{\mathrm{U_{coll}^{\otimes N_g}}}&\ghost{\mathrm{U_{str}}}&\ghost{\mathrm{BC_{L_1 \cap L_2 \cap L_3}}}&\ghost{\mathrm{BC_{L_1}}}&\qw&\ghost{\mathrm{BC_{L_2}}}&\qw&\ghost{\mathrm{BC_{L_3}}}&\qw&\qw\\
\nghost{{m}_{0} :  } & \lstick{\ket{+}_{m_0}} & \qw & \qw & \qw & \gate{\mathrm{X}} & \ctrl{-1} & \qw & \ctrl{-1} & \gate{\mathrm{X}} & \ctrl{-1} & \qw &\qw \\
\nghost{{m}_{1} :  } & \lstick{\ket{+}_{m_1}} & \qw & \qw & \qw & \gate{\mathrm{X}} & \ctrl{-2} & \gate{\mathrm{X}} & \ctrl{-2} & \gate{\mathrm{X}} & \ctrl{-2} & \gate{\mathrm{X}} &\qw \\
  \\}}

%% file: circ-mp-mhwa-schematic.tex
\centering
\scalebox{0.85}{
\Qcircuit @C=1.0em @R=0.2em @!R { \\
\nghost{{v}_{(0, 0)} :  } & \lstick{{v}_{(0, 0)} } & \qw & \multigate{4}{\mathrm{QLGA}}  & \qw & \qw & \qw & \qw & \qw & \qw & \multigate{4}{\mathrm{QLGA}} & \qw & \qw &\qw\\
\nghost{{v}_{(0, 1)} :  } & \lstick{{v}_{(0, 1)} } & \qw & \ghost{\mathrm{QLGA}} & \ctrl{4} & \ctrl{5} & \ctrl{6} & \qw & \qw & \qw & \ghost{\mathrm{QLGA}} & \ctrl{4} & \qw & \qw \\
\nghost{{v}_{(0, 2)} :  } & \lstick{{v}_{(0, 2)} } & \qw & \ghost{\mathrm{QLGA}} & \qw & \qw & \qw & \ctrl{3} & \ctrl{4} & \ctrl{5} & \ghost{\mathrm{QLGA}} & \qw & \ctrl{3} & \qw \\
\nghost{{v}} :  & \lstick{\cdots }\\
\nghost{{v}_{(N_g - 1, N_v - 1)} :  } & \lstick{{v}_{(N_g - 1, N_v - 1)} }  & \qw & \ghost{\mathrm{QLGA}} & \qw & \qw & \qw & \qw & \qw & \qw & \ghost{\mathrm{QLGA}} & \qw & \qw &\qw\\
\nghost{{d}_{0} :  } & \lstick{\ket{0}_{d_0}} & \qw & \multigate{2}{\mathrm{QFT}} & \gate{\mathrm{P}\left(\frac{\uppi}{4}\right)} & \qw & \qw & \gate{\mathrm{P}\left(\frac{\uppi}{4}\right)} & \qw & \qw & \qw & \multigate{2}{\mathrm{P}} & \multigate{2}{\mathrm{P}} & \multigate{2}{\mathrm{QFT}^\dagger}\\
\nghost{{d}_{1} :  } & \lstick{\ket{0}_{d_1}} & \qw & \ghost{\mathrm{QFT}} & \qw & \gate{\mathrm{P}\left(\frac{\uppi}{2}\right)} & \qw & \qw & \gate{\mathrm{P}\left(\frac{\uppi}{2}\right)} & \qw & \qw & \ghost{\mathrm{P}} & \ghost{\mathrm{P}} & \ghost{\mathrm{QFT}^\dagger} \\
\nghost{{d}_{2} :  } & \lstick{\ket{0}_{d_2}} & \qw & \ghost{\mathrm{QFT}} & \qw & \qw & \gate{\mathrm{P}\left(\uppi\right)} & \qw & \qw & \gate{\mathrm{P}\left(\uppi\right)} & \qw  & \ghost{\mathrm{P}} & \ghost{\mathrm{P}} & \ghost{\mathrm{QFT}^\dagger} \\
\\}}

%% file: 4-qae-qmf.tex
\section{Mean Estimation and Minimum Finding \label{sec:mp-4-qaeqmf}}

This section describes how established quantum algorithms
can be used to query the ensemble average LGA quantum state
given in \Cref{eq:mp-3-quantum-state-after-acc} to find the optimal lattice configuration.
\Cref{subsec:mp-4-1-amplitude-projection} describes how we can efficiently transform 
the quantum state into a form that is suitable for the usage of the \dhalg~algorithm,
which is covered in \Cref{subsec:mp-4-1-minimum-finding}.
For applications to the QLBM, the amplitude accumulation
method described by \citet{schalkers2024momentum} can be applied
to obtain the appropriate quantum state without the techniques described in this section.

\subsection{Amplitude Mapping and Mean Estimation \label{subsec:mp-4-1-amplitude-projection}}

The problem of estimating the mean of a (Monte Carlo) quantum algorithm $\mathcal{A}$
 has been studied extensively in the literature.
\citet{heinrich2002quantum} and \citet{brassard2011optimal}
introduced two such asymptotically
optimal algorithms that obtain the typical quadratic speedup associated with
"Grover-like" routines.
\citet{montanaro2015quantum} later described how
these methods can yield this quadratic advantage in tandem
with quantum walks and derived rigorous error bounds by making use the QAE routine.
To make use of these methods, we need to transform the state obtained after
the application of the QLGA algorithm from \Cref{eq:mp-3-quantum-state-after-acc}
into a form that supports the use of QAE.
Previous work covers the construction of such oracles for different classes of problems
including encoding the mean reward in reinforcement learning \cite{dunjko2016quantum},
computing the centroid distance in clustering problems \cite{wiebe2014quantum, kerenidis2019q},
selecting the best arm of a multi-arm bandit in bandit optimization \cite{wang2021quantum},
and topology optimization via finite element solvers in structural mechanics \cite{holscher2025end}.
Such oracles can often be expressed in the form given in \Cref{eq:mp-2-oracle-form}.
Adapting this general form to our LGA setting, we formulate a unitary that
acts on the data and coin registers as

\begin{equation}
    U_{\mathrm{M}}\ket{f(L_j, \Omega)}_{\mathrm{D}}\ket{0}_{\mathrm{C}} = \ket{f(L_j, \Omega)}_{\mathrm{D}}\left( \sqrt{\phi(f)} \ket{1}_{\mathrm{C}} + \sqrt{1-\phi(f)}\ket{0}_{\mathrm{C}} \right),
\end{equation}

while leaving all other registers unaffected.
The purpose of this operation is to map the superposed values
of the $D$ register onto the coin qubit $C$ such that QAE
can then estimate this amplitude.
\footnote{
We note that the usage of amplitude mapping is only required for
computational basis state encodings -- amplitude-based encodings
used widely in QLBM algorithms
may use the techniques outlined in \cite{schalkers2024momentum} and
\cite{georgescu2025fully} as an efficient alternative.
}
We denote this mapping as a function $\phi$,
properties of which warrant some careful consideration.
Specifically, we require that $\phi$ is monotone on the
support imposed by the data register, such that the
minimum finding step can be applied on the QAE of $\phi$ is consistent with $f$.
Second, we require that the implementation of $U_{\mathrm{M}}$ is efficient
in the size of the data register.
Finally, we require that $\phi$ imposes an amplitude that is proportional
to the ensemble average value of $D$ onto the $\ket{1}$ state of the coin qubit,
such that the resulting state matches the goal formulated in \Cref{eq:mp-1-1-problem}.
In the remainder of this section,
we provide two methods implementing such a unitary operation:
a weighted rotation method and a linear comparison method.

\paragraph{Weighted Rotation Mapping.}
The weighted rotation map is implemented by means for a series
of single-controlled $\mathrm{R_Y}(\theta)$ gates,
where the parameter $\theta$ is chosen as the weight of the contributing bit,
from least to most significant.
This operation uses the fact that
$\mathrm{R_Y}(\theta_0)\mathrm{R_Y}(\theta_1)=\mathrm{R_Y}(\theta_0 + \theta_1)$
maps the binary representation of a number $x$ as 

\begin{equation}
    \begin{split}
        \prod_{j=0}^{N_{q,acc} - 1} \mathrm{R_Y}(\alpha x_j2^j) & = \mathrm{R_Y}\left( \sum_{j=0}^{N_{q,acc} - 1} \alpha x_j2^j  \right)\\
        & = \mathrm{R_Y}\left( \alpha x \right),
    \end{split}
\end{equation}

where $x_j$ denotes the state of the $j^{\text{th}}$ qubit of $x$.
Acting on the $\ket{0}$ state, this operation sets the amplitude of the $\ket{1}$ state
to $\sin (\alpha f(L, \Omega)/2)$.
We can enforce the monotonicity of this mapping by setting the value of the free
parameter $\alpha$ to restrict the domain $\phi$.
We know that, by design, $0 \leq f(L, \Omega) \leq N_{t, acc} \| \Omega \| (q + 1) = F_{\mathrm{max}}$,
and we can leverage this fact by setting

\begin{equation}
    \alpha = \frac{\uppi}{F_{\mathrm{max}}} \implies 0 \leq \alpha\frac{f(L, \Omega)}{2} \leq \frac{\uppi F_{\mathrm{max}}}{2F_{\mathrm{max}}}= \frac{\uppi}{2}.
    \label{eq:mp-4-ry-support}
\end{equation}

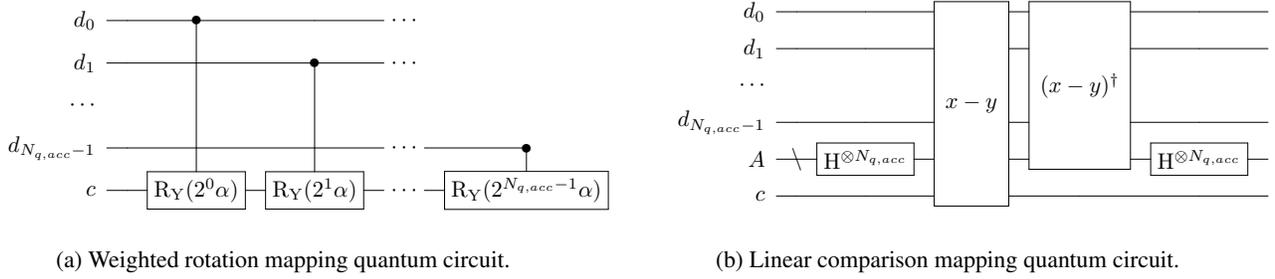
\begin{figure}
    \centering
    \hfill
    \hspace{-1.25cm}
    \subcaptionbox{Weighted rotation mapping quantum circuit.\label{fig:mp-4-amplitude-mapping-ry}}{\input{circ-mp-amplitude-mapping-ry.tex}}
    \hfill
    \subcaptionbox{Linear comparison mapping quantum circuit.\label{fig:mp-4-amplitude-mapping-comp}}{\input{circ-mp-amplitude-mapping-comp.tex}}
    \hfill
    \caption{Amplitude mapping quantum circuits.}
\end{figure}

The relation in \Cref{eq:mp-4-ry-support} limits the support of the sinusoidal $\phi$ function
to $[0, \uppi/2]$, which is an interval in which both the amplitude $\sin$
and its implied probability $\sin^2$ are monotonically increasing.
This circuit requires precisely $N_{q,acc}$ single-controlled
$\mathrm{R_Y}$ gates, implemented in series as shown in \Cref{fig:mp-4-amplitude-mapping-ry}.

\paragraph{Exact Linear Mapping.} A consequence of the sinusoidal shape
of the weighted rotation mapping is that not all regions of the landscape are
equally distinguishable.
This nonlinearity may be beneficial if configurations have values
that are in different regions of the landscape, however, it
is symmetrically a disadvantage when values are in an area of the
$\sin$ function with a shallow slope.
In such instances, a more consistent mapping may become preferable.

The exact linear mapping can be implemented as follows.
In an ancillary register $AM$, of the same number of qubits
as the data accumulation register, we instantiate the uniform superposition
$\ket{u}_\mathrm{AM} = \sum_{j} 1/\sqrt{N_{q,acc}} \ket{j}$.
Next, we use a quantum comparison operation, such as the one
described by \citet{gidney2017factoring},
to compare the values encoded in the data register to $\ket{u}$.
This operation flips the state of the coin qubit for all states $j$
less than the value of the data register.
This is equivalent to $\phi(f) = \sum_xf_x(\cdot)/2^{N_{q,acc}}$.
This form satisfies the monotonicity requirement, and gives an exact linear
comparison between different data register values, per marker.
A practical advantage of this method is that comparator operations
are extensively used in QLGA \cite{georgescu2025fully} and QLBM \cite{schalkers2024efficient}
algorithms for the efficient imposition of boundary conditions,
and implementations are readily available.
The complexity of this step is dominated by the comparison operation,
which is quadratic in the size of the data accumulation register.
Following the comparison and mapping onto the coin qubit, the state
of the ancilla qubits can be reset by undoing the comparison.

Following the application of either amplitude mapping method,
our algorithm proceeds by executing the QAE routine to extract
the amplitudes of the coin register into the estimation register.
Importantly, neither the mapping, nor the QAE steps
need to be controlled on the marker.
Since the parallel QLGA/QLBM algorithm contains blocks that are controlled on the marker,
the application of the Grover iterate $\mathcal{A}_{\mathrm{QLGA}}\mathrm{S}_0\mathcal{A}^\dagger_{\mathrm{QLGA}}S_T$
is itself block-diagonal in the marker states, which implies that
the QAE block acts in parallel for all lattice configurations.
While the complexity of the QAE block itself does depend on the
cost of $\mathcal{A}_{\mathrm{QLGA}}$, the number of queries is
consistent with the typical complexity of $\mathcal{O}(\epsilon^{-1})$ and
no further terms are incurred.

The quantum state obtained after application the amplitude mapping
and the QAE steps is given by

\begin{equation}
    \ket{\psi_{\mathrm{QAE}}} = \sum_{j} \ket{\psi_{L_j}}_{\mathrm{B}}\ket{f(L_j, \Omega)}_{\mathrm{D}}\ket{j}_{\mathrm{M}}\left(  \sqrt{\phi(f_j)} \ket{1}_{\mathrm{C}} + \sqrt{1-\phi(f_j)}\ket{0}_{\mathrm{C}} \right)\ket{\hat{\phi}_j}_{\mathrm{E}}\ket{0}_{\mathrm{G}},
    \label{eq:mp-4-quantum-state-after-qae}
\end{equation}

with $\hat{\phi}$ the estimated parameter such that
$\sin^2{\left(\uppi \hat{\phi}_j / 2^e\right)} \approx \phi(f(L_j, \Omega))$.
The final step of our method uses iterative Grover search
over the estimation register to find the lattice configuration with
the minimum estimated mean QoI.

\subsection{Minimum Finding and Analysis \label{subsec:mp-4-1-minimum-finding}}

The algorithmic steps described thus far
can be regarded as  part of the state preparation procedure that assembles the state

\begin{equation}
    \ket{\psi_{\mathrm{PRE}}} = \frac{1}{\sqrt{\| \mathcal{L} \|}}\sum_{j=0}^{\mathcal{L}} \ket{j}_{\mathrm{M}}\ket{\hat{\phi}_j}_\mathrm{E}\ket{0}_\mathrm{G},
    \label{eq:mp-4-quantum-state-after-qae-simplified}
\end{equation}

with all other registers acting as ancilla space
for the computation of the mean QoIs of each lattice configuration.
The final step of our algorithm consists of running a minimum
finding routine over the register $E$ to find the
minimum estimated value and the marker associated with it.
For this purpose, we use the \dhalg~algorithm \cite{durr1996quantum},
which combines Grover iterations with exponential search.
Intuitively, the \dhalg~routine consists of tracking a
threshold $\tau$ that is used to partition $\ket{\psi_\mathrm{PRE}}$ into
"good" states that fall below the threshold
and "bad" states that exceed $\tau$ as

\begin{equation}
    \ket{\psi_{\mathrm{DH}}} = \frac{1}{\sqrt{\| \mathcal{L} \|}} \sum_{j=0}^{\| \mathcal{L} \| - 1} \ket{j}_{\mathrm{M}}\ket{\hat{\phi}_j}_\mathrm{E}\ket{\hat{\phi_j} < \tau}_\mathrm{G},
    \label{eq:mp-4-quantum-state-after-qae-simplified-threshold}
\end{equation}

which follows exactly the typical
Grover partition form given in \Cref{eq:mp-2-oracle-form}.
This operation only requires one additional comparator operation between the
estimated value and the threshold $\tau$, which can be reversibly implemented
without the use of any ancillary register \cite{schalkers2024efficient}.
Within the Grover search framework, the QLGA $\to$ amplitude mapping $\to$ QAE $\to$ threshold comparison,
followed by the application of a single $\mathrm{Z}$ gate on the $G$ qubit,
can be regarded as the implementation of a phase query gate as described in \Cref{subsubsec:mp-2-3-1-grover}.
By amplifying the amplitudes of the "good" states
that fall below the threshold, the \dhalg~routine
can then measure the outcome of the $E$ register and update
the threshold if a lower value is measured.
By selecting appropriate
update parameters, the typical quadratic Grover search
advantage is preserved, up to a multiplicative constant \cite{durr1996quantum}.
Tying together the query complexity of the minimum finding loop and the
one- and two-qubit gate cost of the parallel QLGA primitives,
we obtain a computational complexity of

\begin{equation}
    \mathcal{O} \left( \underbrace{\sqrt{\| \mathcal{L} \|}}_{\text{\dhalg}} \cdot \overbrace{\frac{1}{\epsilon}}^{\text{QAE}} \cdot N_t \cdot \left(\underbrace{2^q + \log_2 N_g}_{\text{QLGA}} + \overbrace{\log_2^2 \left( N_{t, acc} \cdot \| \Omega \| \cdot (q + 1) \right)}^{\text{Accumulation, Mapping, Comparison}} + \underbrace{N_{bc, L_{\text{max}}} \cdot \| \mathcal{L} \| \cdot \log^2 \| \mathcal{L} \|}_{\text{Parallel Semantics}}\right) \right).
    \label{eq:mp-4-cost-estimate-full}
\end{equation}

\Cref{eq:mp-4-cost-estimate-full} gives a quadratic improvement in the size of the lattice configuration
and in the additive error estimate.
The estimate assumes that initial conditions can be implemented with a cost
that is of the same order as boundary condition imposition,
and accumulation terms absorbs the cost of the amplitude mapping step.
The cost of the \dhalg~comparison and that of the amplitude mapping step
are both absorbed by the accumulation term, which is preformed for multiple time steps.
The QLBM analog would alter the cost of collision and parallel semantics
depending on the implementation.
While the cost of the collision operator varies significantly depending
on the target equation and approximation, the parallel boundary condition imposition
can inherit the polylogarithmic scaling in
the grid size as described in \cite{schalkers2024efficient} and \cite{georgescu2025fully}.
This analysis does not account for the advantage gained by exploiting configuration overlap,
as it is a highly application-dependent.

\paragraph{Error reduction techniques.}
To ensure that the \dhalg~algorithm bounds and precision
are consistent with the theoretical formulation, we must ensure
that the errors incurred by the previous steps are appropriately bounded.
In particular we are concerned with whether the 
the QAE approximation error $\epsilon$ affects the outcome of the minimum
finding routine.
For an amplitude $\phi$ and a positive number of iterations $l$,
the canonical QAE is known to output an estimate
$\hat{\phi}$ that satisfies 

\begin{equation}
    |\phi - \hat{\phi}| \leq 2\uppi\frac{\sqrt{\phi(1-\phi)}}{l} + \left( \frac{\uppi}{l} \right)^2
    \label{eq:mp-4-error-bound}
\end{equation}

with probability at least $8/\uppi^2$ \cite{brassard2000quantum,wiebe2014quantum}.
Furthermore, we consider the gap

\begin{equation}
    \Delta = \mathrm{min}_{L\neq L^\star} \left[ \phi(f(L, \Omega)) - \phi(f({L^\star}, \Omega)) \right]
\end{equation}

between the true optimal solution and the closest other configuration.
In our strictly monotone construction of $\phi$,
errors can occur in instances when QAE precision leads
to misclassifications in the oracle $[\hat{\phi} < \tau]$
as a result of $\Delta$ falling in this range.
This is an instance of a \emph{bounded-error oracle} \cite{hoyer2003quantum},
which can be mitigated in several ways.
\citet{hoyer2003quantum} show that interleaving
amplitude amplification invocations with an error-reduction steps
retains the asymptotic scaling of quantum search in instances
where the oracle error is bounded.
The error reduction routine consists of
stochastically checking whether solutions
marked by the oracle are, in fact, true positives, and
performing majority voting on the outcome to "push back"
false positives \cite{hoyer2003quantum}.
\citet{wiebe2014quantum} also approach the error reduction
problem with a majority voting based scheme, which uses
superposed copies of the QAE estimate to compute the median
of the parallel QAE runs.
Both of these techniques can be applied to our algorithm
in a straightforward manner.



%% file: circ-mp-amplitude-mapping-ry.tex
\centering
\scalebox{0.85}{
\Qcircuit @C=0.9em @R=0.2em @!R { \\
\nghost{{d_0 :  } } & \lstick{d_0} & \qw & \ctrl{4} & \qw & \qw &  \cdots\\
\nghost{{d_1 :  } } & \lstick{d_1} & \qw  & \qw & \ctrl{3} & \qw &  \cdots\\
\nghost{{v}_{(0, 3)} :  } & \lstick{\cdots } \\
\nghost{{d_{N_{q,acc} - 1} :  } } & \lstick{d_{N_{q,acc} - 1} } & \qw & \qw & \qw &\qw &  \cdots &  & \ctrl{1}\\
\nghost{{c :  } } & \lstick{c} & \qw & \gate{\mathrm{R_Y}(2^{0} \alpha)} & \gate{\mathrm{R_Y}(2^{1}\alpha)}&\qw &  \cdots &  &\gate{\mathrm{R_Y}(2^{N_{q,acc} - 1} \alpha)}\\
\\}}

%% file: circ-mp-amplitude-mapping-comp.tex
\centering
\scalebox{0.85}{
\Qcircuit @C=0.9em @R=0.2em @!R { \\
\nghost{{d_0 :  } } & \lstick{d_0} & \qw & \qw & \multigate{5}{x-y} & \multigate{4}{(x-y)^\dagger} &\qw& \qw\\
\nghost{{d_1 :  } } & \lstick{d_1} & \qw  & \qw & \ghost{x-y} & \ghost{(x-y)^\dagger} & \qw& \qw\\
\nghost{{v}_{(0, 3)} :} & \lstick{\cdots } \\
\nghost{{d_{N_{q,acc} - 1} :  } } & \lstick{d_{N_{q,acc} - 1} } & \qw & \qw & \ghost{x-y} & \ghost{(x-y)^\dagger} & \qw& \qw\\
\nghost{{A} : } & \lstick{A} & \qw{\textbackslash} & \gate{\mathrm{H}^{\otimes N_{q,acc}}} & \ghost{x-y} & \ghost{(x-y)^\dagger} & \gate{\mathrm{H}^{\otimes N_{q,acc}}} & \qw\\
\nghost{{c :  } } & \lstick{c} & \qw & \qw & \ghost{x-y}& \qw& \qw& \qw\\
\\}}

%% file: 6-conclusion.tex
\section{Conclusion \label{sec:mp-7-conclusion}}

This work highlighted the connection between Quantum CFD algorithms
and Optimization techniques in a way that circumvents
the requirement of flow field measurement altogether.
We detailed how simple modifications to initial condition calculations
and boundary condition imposition enable the simulation of
many lattice configurations in parallel.
We showed how shared properties of candidate configurations
can be leveraged to increase the efficiency of the parallel QLGA algorithm.
Finally, we described two techniques to transform the quantum state
into a form that enables algorithms from discrete optimization
and machine learning to apply directly to our problem.
Throughout the paper, we provided gate-level descriptions and computational complexity analyses
of the circuits we developed.
With small modifications, the methods described in this paper
apply to both QLGA and QLBM algorithms.

Research at the intersection of Quantum CFD and Optimization
could follow several directions, three of which we address here.
First, the development of specialized Grover diffusion
operators for QLGA and QLBM routines that are more efficient than
inverting the algorithm at the gate level could provide significant
practical performance improvements.
Second, developing methods for QLBM algorithms
that make use of measurement and dynamic circuits
(and therefore violate the coherence assumptions we rely on in our analysis)
would improve the applicability of the discussed methods.
Finally, an error analysis that links physical properties of the system,
such as the Reynolds number, to the accuracy and reliability of the estimation
and minimum finding blocks would provide deeper insight into
the class of problems that Quantum CFD could help solve in the future.

%% file: 7-ack.tex
\section{Acknowledgements\label{sec:ack}}

We gratefully acknowledge support from the joint research program
\emph{Quantum Computational Fluid Dynamics} by Fujitsu Limited and Delft
University of Technology, co-funded by the Netherlands Enterprise
Agency under project number PPS23-3-03596728.

%% file: main.bbl
\begin{thebibliography}{74}
\providecommand{\natexlab}[1]{#1}
\providecommand{\url}[1]{\texttt{#1}}
\expandafter\ifx\csname urlstyle\endcsname\relax
  \providecommand{\doi}[1]{doi: #1}\else
  \providecommand{\doi}{doi: \begingroup \urlstyle{rm}\Url}\fi

\bibitem[Aaronson and Rall(2020)]{aaronson2020quantum}
Scott Aaronson and Patrick Rall.
\newblock Quantum approximate counting, simplified.
\newblock In \emph{Symposium on simplicity in algorithms}, pages 24--32. SIAM, 2020.

\bibitem[Barenco et~al.(1995)Barenco, Bennett, Cleve, DiVincenzo, Margolus, Shor, Sleator, Smolin, and Weinfurter]{barenco1995elementary}
Adriano Barenco, Charles~H Bennett, Richard Cleve, David~P DiVincenzo, Norman Margolus, Peter Shor, Tycho Sleator, John~A Smolin, and Harald Weinfurter.
\newblock Elementary gates for quantum computation.
\newblock \emph{Physical review A}, 52\penalty0 (5):\penalty0 3457, 1995.

\bibitem[Bhatnagar et~al.(1954)Bhatnagar, Gross, and Krook]{bhatnagar1954model}
Prabhu~Lal Bhatnagar, Eugene~P Gross, and Max Krook.
\newblock A model for collision processes in gases. i. small amplitude processes in charged and neutral one-component systems.
\newblock \emph{Physical review}, 94\penalty0 (3):\penalty0 511, 1954.

\bibitem[Boghosian and Taylor(1997)]{boghosian1997quantum}
Bruce~M Boghosian and Washington Taylor.
\newblock Quantum lattice-gas models for the many-body schr{\"o}dinger equation.
\newblock \emph{International Journal of Modern Physics C}, 8\penalty0 (04):\penalty0 705--716, 1997.

\bibitem[Boghosian and Taylor~IV(1998)]{boghosian1998simulating}
Bruce~M Boghosian and Washington Taylor~IV.
\newblock Simulating quantum mechanics on a quantum computer.
\newblock \emph{Physica D: Nonlinear Phenomena}, 120\penalty0 (1-2):\penalty0 30--42, 1998.

\bibitem[Boyer et~al.(1998)Boyer, Brassard, H{\o}yer, and Tapp]{boyer1998tight}
Michel Boyer, Gilles Brassard, Peter H{\o}yer, and Alain Tapp.
\newblock Tight bounds on quantum searching.
\newblock \emph{Fortschritte der Physik: Progress of Physics}, 46\penalty0 (4-5):\penalty0 493--505, 1998.

\bibitem[Brassard et~al.(2000)Brassard, Hoyer, Mosca, and Tapp]{brassard2000quantum}
Gilles Brassard, Peter Hoyer, Michele Mosca, and Alain Tapp.
\newblock Quantum amplitude amplification and estimation.
\newblock \emph{arXiv preprint quant-ph/0005055}, 2000.

\bibitem[Brassard et~al.(2011)Brassard, Dupuis, Gambs, and Tapp]{brassard2011optimal}
Gilles Brassard, Frederic Dupuis, Sebastien Gambs, and Alain Tapp.
\newblock An optimal quantum algorithm to approximate the mean and its application for approximating the median of a set of points over an arbitrary distance.
\newblock \emph{arXiv preprint arXiv:1106.4267}, 2011.

\bibitem[Budinski(2021)]{budinski2021quantum}
Ljubomir Budinski.
\newblock Quantum algorithm for the advection--diffusion equation simulated with the lattice boltzmann method.
\newblock \emph{Quantum Information Processing}, 20\penalty0 (2):\penalty0 57, 2021.

\bibitem[Budinski(2022)]{ljubomir2022quantum}
Ljubomir Budinski.
\newblock Quantum algorithm for the navier--stokes equations by using the streamfunction-vorticity formulation and the lattice boltzmann method.
\newblock \emph{International Journal of Quantum Information}, 20\penalty0 (02):\penalty0 2150039, 2022.

\bibitem[Childs and Wiebe(2012)]{childs2012hamiltonian}
Andrew~M Childs and Nathan Wiebe.
\newblock Hamiltonian simulation using linear combinations of unitary operations.
\newblock \emph{arXiv preprint arXiv:1202.5822}, 2012.

\bibitem[Demirdjian et~al.(2022)Demirdjian, Gunlycke, Reynolds, Doyle, and Tafur]{demirdjian2022variational}
Reuben Demirdjian, Daniel Gunlycke, Carolyn~A Reynolds, James~D Doyle, and Sergio Tafur.
\newblock Variational quantum solutions to the advection--diffusion equation for applications in fluid dynamics.
\newblock \emph{Quantum Information Processing}, 21\penalty0 (9):\penalty0 322, 2022.

\bibitem[Draper(2000)]{draper2000addition}
Thomas~G Draper.
\newblock Addition on a quantum computer.
\newblock \emph{arXiv preprint quant-ph/0008033}, 2000.

\bibitem[Dunjko et~al.(2016)Dunjko, Taylor, and Briegel]{dunjko2016quantum}
Vedran Dunjko, Jacob~M Taylor, and Hans~J Briegel.
\newblock Quantum-enhanced machine learning.
\newblock \emph{Physical review letters}, 117\penalty0 (13):\penalty0 130501, 2016.

\bibitem[Durr and Hoyer(1996)]{durr1996quantum}
Christoph Durr and Peter Hoyer.
\newblock A quantum algorithm for finding the minimum.
\newblock \emph{arXiv preprint quant-ph/9607014}, 1996.

\bibitem[Esmaeilifar et~al.(2024)Esmaeilifar, Ahn, and Myong]{esmaeilifar2024quantum}
Esmaeil Esmaeilifar, Doyeol Ahn, and Rho~Shin Myong.
\newblock Quantum algorithm for nonlinear burgers' equation for high-speed compressible flows.
\newblock \emph{Physics of Fluids}, 36\penalty0 (10), 2024.

\bibitem[Farrelly(2020)]{farrelly2020review}
Terry Farrelly.
\newblock A review of quantum cellular automata.
\newblock \emph{Quantum}, 4:\penalty0 368, 2020.

\bibitem[Fonio et~al.(2025{\natexlab{a}})Fonio, Sagaut, Budinski, and Lahtinen]{fonio2025adaptive}
Niccol{\`o} Fonio, Pierre Sagaut, Ljubomir Budinski, and Valtteri Lahtinen.
\newblock Adaptive lattice-gas algorithm: Classical and quantum implementations.
\newblock \emph{Physical Review E}, 112\penalty0 (3):\penalty0 035302, 2025{\natexlab{a}}.

\bibitem[Fonio et~al.(2025{\natexlab{b}})Fonio, Sagaut, and {Di Molfetta}]{fonio2023quantum}
Niccolò Fonio, Pierre Sagaut, and Giuseppe {Di Molfetta}.
\newblock Quantum collision circuit, quantum invariants and quantum phase estimation procedure for fluid dynamic lattice gas automata.
\newblock \emph{Computers \& Fluids}, 299:\penalty0 106688, 2025{\natexlab{b}}.
\newblock ISSN 0045-7930.

\bibitem[Frisch et~al.(1986)Frisch, Hasslacher, and Pomeau]{frisch1986lattice}
Uriel Frisch, Brosl Hasslacher, and Yves Pomeau.
\newblock Lattice-gas automata for the navier-stokes equation.
\newblock \emph{Physical review letters}, 56\penalty0 (14):\penalty0 1505, 1986.

\bibitem[Gaitan(2020)]{gaitan2020finding}
Frank Gaitan.
\newblock Finding flows of a navier--stokes fluid through quantum computing.
\newblock \emph{npj Quantum Information}, 6\penalty0 (1):\penalty0 61, 2020.

\bibitem[Georgescu et~al.(2025{\natexlab{a}})Georgescu, Schalkers, and M{\"o}ller]{georgescu2025fully}
C{\u{a}}lin~A Georgescu, Merel~A Schalkers, and Matthias M{\"o}ller.
\newblock Fully quantum lattice gas automata building blocks for computational basis state encodings.
\newblock \emph{arXiv preprint arXiv:2506.12662}, 2025{\natexlab{a}}.

\bibitem[Georgescu et~al.(2025{\natexlab{b}})Georgescu, Schalkers, and M{\"o}ller]{georgescu2025qlbm}
C{\u{a}}lin~A Georgescu, Merel~A Schalkers, and Matthias M{\"o}ller.
\newblock qlbm--a quantum lattice boltzmann software framework.
\newblock \emph{Computer Physics Communications}, page 109699, 2025{\natexlab{b}}.

\bibitem[Gidney(2017)]{gidney2017factoring}
Craig Gidney.
\newblock Factoring with n+ 2 clean qubits and n-1 dirty qubits.
\newblock \emph{arXiv preprint arXiv:1706.07884}, 2017.

\bibitem[Givi et~al.(2020)Givi, Daley, Mavriplis, and Malik]{givi2020quantum}
Peyman Givi, Andrew~J Daley, Dimitri Mavriplis, and Mujeeb Malik.
\newblock Quantum speedup for aeroscience and engineering.
\newblock \emph{AIAA Journal}, 58\penalty0 (8):\penalty0 3715--3727, 2020.

\bibitem[Grinko et~al.(2021)Grinko, Gacon, Zoufal, and Woerner]{grinko2021iterative}
Dmitry Grinko, Julien Gacon, Christa Zoufal, and Stefan Woerner.
\newblock Iterative quantum amplitude estimation.
\newblock \emph{npj Quantum Information}, 7\penalty0 (1):\penalty0 52, 2021.

\bibitem[Grover(1996)]{grover1996fast}
Lov~K Grover.
\newblock A fast quantum mechanical algorithm for database search.
\newblock In \emph{Proceedings of the twenty-eighth annual ACM symposium on Theory of computing}, pages 212--219, 1996.

\bibitem[Grover(1997)]{grover1997quantum}
Lov~K Grover.
\newblock Quantum mechanics helps in searching for a needle in a haystack.
\newblock \emph{Physical review letters}, 79\penalty0 (2):\penalty0 325, 1997.

\bibitem[Hardy et~al.(1973)Hardy, Pomeau, and De~Pazzis]{hardy1973time}
J~Hardy, Yves Pomeau, and O~De~Pazzis.
\newblock Time evolution of a two-dimensional model system. i. invariant states and time correlation functions.
\newblock \emph{Journal of Mathematical Physics}, 14\penalty0 (12):\penalty0 1746--1759, 1973.

\bibitem[Heinrich(2002)]{heinrich2002quantum}
Stefan Heinrich.
\newblock Quantum summation with an application to integration.
\newblock \emph{Journal of Complexity}, 18\penalty0 (1):\penalty0 1--50, 2002.

\bibitem[Herbert et~al.(2024)Herbert, Sorci, and Tang]{herbert2024almost}
Steven Herbert, Julien Sorci, and Yao Tang.
\newblock Almost-optimal computational-basis-state transpositions.
\newblock \emph{Physical Review A}, 110\penalty0 (1):\penalty0 012437, 2024.

\bibitem[H{\"o}lscher et~al.(2025)H{\"o}lscher, Ahrend, Karch, L'Estocq, Andreu, Stollenwerk, Wilhelm, and Kowalski]{holscher2025end}
Leonhard H{\"o}lscher, Oliver Ahrend, Lukas Karch, Carlotta L'Estocq, Marc~Marfany Andreu, Tobias Stollenwerk, Frank~K Wilhelm, and Julia Kowalski.
\newblock End-to-end quantum algorithm for topology optimization in structural mechanics.
\newblock \emph{arXiv preprint arXiv:2510.07280}, 2025.

\bibitem[H{\o}yer et~al.(2003)H{\o}yer, Mosca, and De~Wolf]{hoyer2003quantum}
Peter H{\o}yer, Michele Mosca, and Ronald De~Wolf.
\newblock Quantum search on bounded-error inputs.
\newblock In \emph{International Colloquium on Automata, Languages, and Programming}, pages 291--299. Springer, 2003.

\bibitem[Itani(2025)]{itani2025towards}
Wael Itani.
\newblock \emph{Towards a Quantum Algorithm for Lattice Boltzmann (QALB) Simulation With a Nonlinear Collision Term}.
\newblock New York University Tandon School of Engineering, 2025.

\bibitem[Itani and Succi(2022)]{itani2022analysis}
Wael Itani and Sauro Succi.
\newblock Analysis of carleman linearization of lattice boltzmann.
\newblock \emph{Fluids}, 7\penalty0 (1):\penalty0 24, 2022.

\bibitem[Itani et~al.(2023)Itani, Sreenivasan, and Succi]{itani2023qalb}
Wael Itani, Katepalli~R. Sreenivasan, and Sauro Succi.
\newblock Quantum algorithm for lattice boltzmann (qalb) simulation of incompressible fluids with a nonlinear collision term, 2023.
\newblock URL \url{https://arxiv.org/abs/2304.05915}.

\bibitem[Jawetz et~al.(2025)Jawetz, Song, Bryngelson, and Alexeev]{jawetz2025quantum}
Christopher~L Jawetz, Zhixin Song, Spencer~H Bryngelson, and Alexander Alexeev.
\newblock Quantum lattice boltzmann algorithm for heat transfer with phase change.
\newblock \emph{arXiv preprint arXiv:2509.21630}, 2025.

\bibitem[Kerenidis et~al.(2019)Kerenidis, Landman, Luongo, and Prakash]{kerenidis2019q}
Iordanis Kerenidis, Jonas Landman, Alessandro Luongo, and Anupam Prakash.
\newblock q-means: A quantum algorithm for unsupervised machine learning.
\newblock \emph{Advances in neural information processing systems}, 32, 2019.

\bibitem[Kocherla et~al.(2024)Kocherla, Song, Chrit, Gard, Dumitrescu, Alexeev, and Bryngelson]{kocherla2024fully}
Sriharsha Kocherla, Zhixin Song, Fatima~Ezahra Chrit, Bryan Gard, Eugene~F Dumitrescu, Alexander Alexeev, and Spencer~H Bryngelson.
\newblock Fully quantum algorithm for mesoscale fluid simulations with application to partial differential equations.
\newblock \emph{AVS Quantum Science}, 6\penalty0 (3), 2024.

\bibitem[Kr{\"u}ger et~al.(2017)Kr{\"u}ger, Kusumaatmaja, Kuzmin, Shardt, Silva, and Viggen]{kruger2017lattice}
Timm Kr{\"u}ger, Halim Kusumaatmaja, Alexandr Kuzmin, Orest Shardt, Goncalo Silva, and Erlend~Magnus Viggen.
\newblock The lattice boltzmann method.
\newblock \emph{Springer International Publishing}, 10\penalty0 (978-3):\penalty0 4--15, 2017.

\bibitem[Kyriienko et~al.(2021)Kyriienko, Paine, and Elfving]{kyriienko2021solving}
Oleksandr Kyriienko, Annie~E Paine, and Vincent~E Elfving.
\newblock Solving nonlinear differential equations with differentiable quantum circuits.
\newblock \emph{Physical Review A}, 103\penalty0 (5):\penalty0 052416, 2021.

\bibitem[L{\u{a}}c{\u{a}}tu{\c{s}} and M{\"o}ller(2025)]{luacuatucs2025surrogate}
Monica L{\u{a}}c{\u{a}}tu{\c{s}} and Matthias M{\"o}ller.
\newblock Surrogate quantum circuit design for the lattice boltzmann collision operator.
\newblock \emph{arXiv preprint arXiv:2507.12256}, 2025.

\bibitem[Love(2019)]{love2019quantum}
Peter Love.
\newblock On quantum extensions of hydrodynamic lattice gas automata.
\newblock \emph{Condensed Matter}, 4\penalty0 (2):\penalty0 48, 2019.

\bibitem[Mani and Dorgan(2023)]{mani2023perspective}
Mori Mani and Andrew~J Dorgan.
\newblock A perspective on the state of aerospace computational fluid dynamics technology.
\newblock \emph{Annual Review of Fluid Mechanics}, 55\penalty0 (1):\penalty0 431--457, 2023.

\bibitem[Meyer(1996)]{meyer1996quantum}
David~A Meyer.
\newblock From quantum cellular automata to quantum lattice gases.
\newblock \emph{Journal of Statistical Physics}, 85:\penalty0 551--574, 1996.

\bibitem[Meyer(1997)]{meyer1997quantumb}
David~A Meyer.
\newblock Quantum lattice gases and their invariants.
\newblock \emph{International Journal of Modern Physics C}, 8\penalty0 (04):\penalty0 717--735, 1997.

\bibitem[Montanaro(2015)]{montanaro2015quantum}
Ashley Montanaro.
\newblock Quantum speedup of monte carlo methods.
\newblock \emph{Proceedings of the Royal Society A: Mathematical, Physical and Engineering Sciences}, 471\penalty0 (2181):\penalty0 20150301, 2015.

\bibitem[Nagel and L{\"o}we(2025)]{nagel2025quantum}
Aaron Nagel and Johannes L{\"o}we.
\newblock Quantum lattice boltzmann method for multiple time steps without reinitialization for linear advection-diffusion problems.
\newblock \emph{arXiv preprint arXiv:2510.05965}, 2025.

\bibitem[Nayak and Wu(1999)]{nayak1999quantum}
Ashwin Nayak and Felix Wu.
\newblock The quantum query complexity of approximating the median and related statistics.
\newblock In \emph{Proceedings of the thirty-first annual ACM symposium on Theory of computing}, pages 384--393, 1999.

\bibitem[Sanavio and Succi(2024)]{sanavio2024lattice}
Claudio Sanavio and Sauro Succi.
\newblock Lattice boltzmann--carleman quantum algorithm and circuit for fluid flows at moderate reynolds number.
\newblock \emph{AVS Quantum Science}, 6\penalty0 (2), 2024.

\bibitem[Sanavio et~al.(2025)Sanavio, Simon, Ralli, Love, and Succi]{sanavio2025carleman}
Claudio Sanavio, William~A. Simon, Alexis Ralli, Peter Love, and Sauro Succi.
\newblock Carleman-lattice-boltzmann quantum circuit with matrix access oracles, 2025.
\newblock URL \url{https://arxiv.org/abs/2501.02582}.

\bibitem[Schalkers and M{\"o}ller(2024{\natexlab{a}})]{schalkers2024efficient}
Merel~A Schalkers and Matthias M{\"o}ller.
\newblock Efficient and fail-safe quantum algorithm for the transport equation.
\newblock \emph{Journal of Computational Physics}, 502:\penalty0 112816, 2024{\natexlab{a}}.

\bibitem[Schalkers and M{\"o}ller(2024{\natexlab{b}})]{schalkers2024importance}
Merel~A Schalkers and Matthias M{\"o}ller.
\newblock On the importance of data encoding in quantum boltzmann methods.
\newblock \emph{Quantum Information Processing}, 23\penalty0 (1):\penalty0 20, 2024{\natexlab{b}}.

\bibitem[Schalkers and M{\"o}ller(2024{\natexlab{c}})]{schalkers2024momentum}
Merel~A Schalkers and Matthias M{\"o}ller.
\newblock Momentum exchange method for quantum boltzmann methods.
\newblock \emph{Computers \& Fluids}, 285:\penalty0 106453, 2024{\natexlab{c}}.

\bibitem[Schaller(1997)]{schaller1997moore}
Robert~R Schaller.
\newblock Moore's law: past, present and future.
\newblock \emph{IEEE spectrum}, 34\penalty0 (6):\penalty0 52--59, 1997.

\bibitem[Steijl(2020)]{steijl2020quantum}
Rene Steijl.
\newblock Quantum algorithms for nonlinear equations in fluid mechanics.
\newblock \emph{Quantum computing and communications}, 2020.

\bibitem[Succi(2001)]{succi2001lattice}
Sauro Succi.
\newblock \emph{The lattice Boltzmann equation: for fluid dynamics and beyond}.
\newblock Oxford university press, 2001.

\bibitem[Suzuki et~al.(2020)Suzuki, Uno, Raymond, Tanaka, Onodera, and Yamamoto]{suzuki2020amplitude}
Yohichi Suzuki, Shumpei Uno, Rudy Raymond, Tomoki Tanaka, Tamiya Onodera, and Naoki Yamamoto.
\newblock Amplitude estimation without phase estimation: Y. suzuki et al.
\newblock \emph{Quantum Information Processing}, 19\penalty0 (2):\penalty0 75, 2020.

\bibitem[Todorova and Steijl(2020)]{todorova2020quantum}
Blaga~N Todorova and Ren{\'e} Steijl.
\newblock Quantum algorithm for the collisionless boltzmann equation.
\newblock \emph{Journal of Computational Physics}, 409:\penalty0 109347, 2020.

\bibitem[v~Neumann and Burks(1966)]{v1966theory}
John v~Neumann and Arthur~W Burks.
\newblock \emph{Theory of self-reproducing automata}.
\newblock University of Illinois Press Urbana, 1966.

\bibitem[Venegas-Andraca(2012)]{venegas2012quantum}
Salvador~El{\'\i}as Venegas-Andraca.
\newblock Quantum walks: a comprehensive review.
\newblock \emph{Quantum Information Processing}, 11\penalty0 (5):\penalty0 1015--1106, 2012.

\bibitem[Wang et~al.(2025)Wang, Meng, Zhao, and Yang]{wang2025quantum}
Boyuan Wang, Zhaoyuan Meng, Yaomin Zhao, and Yue Yang.
\newblock Quantum lattice boltzmann method for simulating nonlinear fluid dynamics.
\newblock \emph{arXiv preprint arXiv:2502.16568}, 2025.

\bibitem[Wang et~al.(2021)Wang, You, Li, and Childs]{wang2021quantum}
Daochen Wang, Xuchen You, Tongyang Li, and Andrew~M Childs.
\newblock Quantum exploration algorithms for multi-armed bandits.
\newblock In \emph{Proceedings of the AAAI Conference on Artificial Intelligence}, volume~35, pages 10102--10110, 2021.

\bibitem[Wawrzyniak et~al.(2025{\natexlab{a}})Wawrzyniak, Winter, Schmidt, Indinger, Jan{\ss}en, Schramm, and Adams]{wawrzyniak2025linearized}
David Wawrzyniak, Josef Winter, Steffen Schmidt, Thomas Indinger, Christian~F Jan{\ss}en, Uwe Schramm, and Nikolaus~A Adams.
\newblock Linearized quantum lattice-boltzmann method for the advection-diffusion equation using dynamic circuits.
\newblock \emph{Computer Physics Communications}, page 109856, 2025{\natexlab{a}}.

\bibitem[Wawrzyniak et~al.(2025{\natexlab{b}})Wawrzyniak, Winter, Schmidt, Indiniger, Jan{\ss}en, Schramm, and Adams]{wawrzyniak2025dynamic}
David Wawrzyniak, Josef Winter, Steffen Schmidt, Thomas Indiniger, Christian~F Jan{\ss}en, Uwe Schramm, and Nikolaus~A Adams.
\newblock Dynamic circuits for the quantum lattice-boltzmann method.
\newblock \emph{arXiv preprint arXiv:2502.02131}, 2025{\natexlab{b}}.

\bibitem[Wie(2019)]{wie2019simpler}
Chu-Ryang Wie.
\newblock Simpler quantum counting.
\newblock \emph{arXiv preprint arXiv:1907.08119}, 2019.

\bibitem[Wiebe et~al.(2015)Wiebe, Kapoor, and Svore]{wiebe2014quantum}
Nathan Wiebe, Ashish Kapoor, and Krysta~M Svore.
\newblock Quantum nearest-neighbor algorithms for machine learning.
\newblock \emph{Quantum information and computation}, 15\penalty0 (3-4):\penalty0 318--358, 2015.

\bibitem[Wolf-Gladrow(2004)]{wolf2004lattice}
Dieter~A Wolf-Gladrow.
\newblock \emph{Lattice-gas cellular automata and lattice Boltzmann models: an introduction}.
\newblock Springer, 2004.

\bibitem[Yepez(1998{\natexlab{a}})]{yepez1998lattice}
Jeffrey Yepez.
\newblock Lattice-gas quantum computation.
\newblock \emph{International Journal of Modern Physics C}, 9\penalty0 (08):\penalty0 1587--1596, 1998{\natexlab{a}}.

\bibitem[Yepez(1998{\natexlab{b}})]{yepez1998quantum}
Jeffrey Yepez.
\newblock Quantum computation of fluid dynamics.
\newblock In \emph{NASA International Conference on Quantum Computing and Quantum Communications}, pages 34--60. Springer, 1998{\natexlab{b}}.

\bibitem[Yepez(2001)]{yepez2001quantum}
Jeffrey Yepez.
\newblock Quantum lattice-gas model for computational fluid dynamics.
\newblock \emph{Physical Review E}, 63\penalty0 (4):\penalty0 046702, 2001.

\bibitem[Yepez(2002)]{yepez2002quantum}
Jeffrey Yepez.
\newblock Quantum lattice-gas model for the burgers equation.
\newblock \emph{Journal of Statistical Physics}, 107:\penalty0 203--224, 2002.

\bibitem[Zamora et~al.(2025{\natexlab{a}})Zamora, Budinski, Niemim{\"a}ki, and Lahtinen]{zamora2025efficient}
Antonio David~Bastida Zamora, Ljubomir Budinski, Ossi Niemim{\"a}ki, and Valtteri Lahtinen.
\newblock Efficient quantum lattice gas automata.
\newblock \emph{Computers \& Fluids}, 286:\penalty0 106476, 2025{\natexlab{a}}.

\bibitem[Zamora et~al.(2025{\natexlab{b}})Zamora, Budinski, Sagaut, and Lahtinen]{zamora2025float}
Antonio David~Bastida Zamora, Ljubomir Budinski, Pierre Sagaut, and Valtteri Lahtinen.
\newblock Lattice gas automata with floating-point numbers: A connection between molecular dynamics and lattice boltzmann method for quantum computers.
\newblock \emph{Physical Review E}, 112\penalty0 (1):\penalty0 015305, 2025{\natexlab{b}}.

\end{thebibliography}
